\documentclass[journal, onecolumn,12pt]{IEEEtran}
\usepackage{epsfig,makeidx,color}
\usepackage{algpseudocode}
\usepackage{url}
\usepackage{graphicx}
\usepackage{graphics}
\usepackage{caption}
\usepackage{subcaption}
\usepackage{psfrag}
\usepackage{latexsym}
\usepackage{showidx}
\usepackage{latexsym}
\usepackage{amssymb}
\usepackage{amsfonts}
\usepackage{amsmath}
\usepackage{graphicx}
\usepackage{amssymb}
\usepackage{amsfonts}
\usepackage{amsmath}
\usepackage{setspace}
\usepackage{lipsum}
\usepackage[font=scriptsize]{caption}
\ifCLASSINFOpdf
\else
\fi
\newtheorem{Theorem}{Theorem}

\newtheorem{Proposition}{Proposition}
\newtheorem{Example}{Example}
\newtheorem{lem}{Lemma}
\newtheorem{lemma}{Lemma}[section]
\newtheorem{Definition}{Definition}

\newcommand{\rmv}[1]{}
\begin{document}
%%%%%%%%%%%%%%%%%%%%%%%%%%%%%%%%%%%%%%%%%%%%%%%%%%%%%%%%%%%%%%%%%%%%%%%%%%%%%%%%%%%%%%%%%
\title{Phase Precoding for the Compute-and-Forward Protocol}
%%%%%%%%%%%%%%%%%%%%%%%%%%%%%%%%%%%%%%%%%%%%%%%%%%%%%%%%%%%%%%%%%%%%%%%%%%%%%%%%%%%%%%%%%
\author{Amin Sakzad, Emanuele Viterbo, Joseph Jean Boutros, and Yi Hong
\thanks{Amin Sakzad, Emanuele Viterbo, and Yi Hong are with
Department of Electrical and Computer Systems, Faculty of Engineering,
Monash University, Clayton VIC 3800, Australia.
E-mail: $\tt \{amin.sakzad,emanuele.viterbo,yi.hong$\}@$\tt monash.edu$.
Joseph J. Boutros is with Department of Electrical Engineering,
Texas A\&M University at Qatar, Doha, Qatar. Email:
$\tt boutros$@$\tt tamu.edu$.

This work was performed at the Monash Software Defined Telecommunications (SDT) Lab and was supported by the Australian Research Council under Discovery grants ARC DP~$130100103$ and NPRP grant NPRP5-597-2-241 from the Qatar National Research Fund (a member of Qatar Foundation). }

\thanks{A subset of this work was presented in~\cite{sakzad14-1} at ISIT 2014, Honolulu, HI, USA.}}
\maketitle
\pagestyle{empty}
\thispagestyle{empty}
\IEEEpeerreviewmaketitle
%%%%%%%%%%%%%%%%%%%%%%%%%%%%%%%%%%%%%%%%%%%%%%%%%%%%%%%%%%%%%%%%%%%%%%%%%%%%%%%%%%%%%%%%%
\begin{abstract}
The compute-and-forward (CoF) is a relaying protocol, which uses algebraic structured codes to harness the interference and remove the noise in wireless networks. We propose the use of phase precoders at the transmitters of a network, where relays apply CoF strategy. We define the {\em phase precoded computation rate} and show that it is greater than the original computation rate of CoF protocol. We further give a new low-complexity method for finding network equations. We finally show that the proposed precoding scheme increases the degrees-of-freedom (DoF) of CoF protocol. This overcomes the limitations on the DoF of the CoF protocol, recently presented by Niesen and Whiting. Using tools from Diophantine approximation and algebraic geometry, we prove the existence of a phase precoder that approaches the maximum DoF when the number of transmitters tends to infinity.
%Finally, we conduct computer simulations to verify the effectiveness of the proposed phase precoding technique. We further give a new low-complexity method for finding network equations. The gain in computation rate and the higher %DoF expected from theoretical results are confirmed by computer simulations with lattice codes.
\end{abstract}
%%%%%%%%%%%%%%%%%%%%%%%%%%%%%%%%%%%%%%%%%%%%%%%%%%%%%%%%%%%%%%%
\begin{IEEEkeywords}
Compute-and-forward, phase precoding, degrees-of-freedom, lattice codes, Diophantine approximation.
\end{IEEEkeywords}
%%%%%%%%%%%%%%%%%%%%%%%%%%%%%%%%%%%%%%%%%%%%%%%%%%%%%%%%%%%%%%%
\section{Introduction}
%%%%%%%%%%%%%%%%%%%%%%%%%%%%%%%%%%%%%%%%%%%%%%%%%%%%%%%%%%%%%%%
Network coding is a highly efficient technique for exchanging information over relaying networks that are central to the most recent wireless communication systems. The rapid expansion of the application of wireless networks has promoted researchers to deal with more complex channels, which are affected by both fading and interference. In the presence of errors, fading, and interference, new algebraic code designs are needed to improve the poor performance of wireless networks~\cite{Feng,Narayanan07,popovski07, Narayanan10}. The diversity techniques are used to combat channel fading~\cite{Tse_visw}. Different cooperative transmission protocols can be categorized into four principal classes: the amplify-and-forward (AF) scheme~\cite{AF, AF1, AF2}, decode-and-forward (DF) strategy~\cite{DF1,DF2}, compress-and-forward (CF) scheme~\cite{Lim11}, and Compute-and-Forward (CoF) protocol~\cite{Feng,Nazer11}. We only focus on the latest technique which improves the network throughput. In particular, we show that phase precoding (PP) improves the degrees-of-freedom (DoF) of CoF and yields higher coding gains for network equation error rates. PP acts on CoF in a way similar to diversity techniques on channel fading~\cite{Tse_visw}.

Let us briefly recall the CoF protocol with two users and one relay~\cite{Nazer11}: suppose that ${\bf x}_1$ and ${\bf x}_2$ are the transmitted complex lattice codewords from the first and the second user, respectively. The received vector at the relay is $h_1{\bf x}_1+h_2{\bf x}_2+{\bf z}$, where ${\bf z}$ is a Gaussian noise and the components of ${\bf h}=(h_1,~h_2)$ are the complex channel fading coefficients from the first and the second user to the relay, respectively. The task of the relay is to estimate an integer linear combination $a_1{\bf x}_1+a_2{\bf x}_2$ from the received vector. The estimated lattice point $a_1{\bf x}_1+a_2{\bf x}_2$ is still a lattice point because any integer linear combination of lattice points is a lattice point. The quality of such an estimate and consequently the computation rate is controlled by a non-zero coefficient $\alpha$. In particular, the parameter $\alpha$ and the integer vector ${\bf a}=(a_1,~a_2)$ are chosen so that $\alpha{\bf h} \approx {\bf a}$.

The computation rate for CoF was introduced in~\cite{Nazer11}. The design of algebraic lattice codes based on Smith normal form for lattice network coding (LNC) using CoF has been investigated in~\cite{Feng}. In addition, a multiple-input multiple-output (MIMO) CoF (Integer-Forcing (IF)) linear receiver architecture is proposed in~\cite{zhan12} based on the idea of CoF.

To implement the CoF protocol, it is necessary to find an optimal non-singular integer matrix consisting of all the network equation coefficients. The solution to this problem is addressed in~\cite{Feng, Sakzad13-1,Sakzad13-2} using lattice reduction algorithms~\cite{CLLL09,Lagarias90, Minkowski}. Furthermore, Hong {\em et al.} have used CoF and IF to achieve higher rates in cooperative distributed antenna systems~\cite{caire}.

Using results from the theory of Diophantine approximation~\cite{Waldshmidt11}, Niesen and Whiting~\cite{Niesen} have recently shown that the CoF protocol has limited degrees-of-freedom (DoF). Then, they have considered channel state information at the transmitters (CSIT) and proposed ``signal alignment'' to achieve full DoF of an $L\times L$ section of a network, whose relays employ CoF~\cite{Niesen2, Niesen}. Overall, Diophantine approximation delivers a positive and a negative result in~\cite{Niesen}. The results from Diophantine approximation~\cite{B01, BBKM02} have also been used to obtain desired DoF in interference channels using interference alignment~\cite{Mahboubui10,Motahari09}. All of the above mentioned applications of Diophantine approximation in the analysis of telecommunication systems have used convergent parts of the results from metric number theory. The theory of Diophantine approximation is a powerful tool for the analysis of DoF in various communication set-ups. For a survey on recent advances in Diophantine approximation theory, we refer the reader to~\cite{Waldshmidt11} and references therein. A simple introduction on Diophantine approximation theory is given in Appendix~\ref{backDio}.

Khintchine and Dirichlet theorems~\cite{Waldshmidt11} (see Appendix~\ref{backDio}) provide results about the approximation of real numbers by rationals, where rational numbers are algebraic numbers of degree one. Some notions on algebraic numbers and hypersurfaces are given in Appendix~\ref{backDio}. In a seminal paper, Davenport and Schmidt~\cite{DS67} have shown that the approximation of real numbers by algebraic numbers of degree at most two has a faster convergence speed. This convergence speed improvement can also be generalized to approximation of real vectors by algebraic hypersurfaces. This occurs because we enlarge our approximant space from rational numbers to algebraic numbers or hypersurfaces.

This suggests that, in order to have better approximations, we should broaden the approximant space. In particular, the approximation of $\alpha{\bf h}$ by integer vector ${\bf a}$, which is shown in~\cite{Niesen} to be an instance of approximation of real vectors by rational ones, is the main limitation to the DoF in the CoF protocol. We expand the appoximant domain from integers to a  broader set by introducing a phase precoding technique. This results in an improved DoF from the maximum of $1/2$ to almost $1$ for almost every channel realization.

%The main idea behind phase precoding is exactly to make the approximant space for $\alpha{\bf h}$ bigger than just integers and rational vectors. Indeed, the precoder for each transmitter is a complex scalar $e^{i\phi}$, for some %$-\pi/4\leq\phi\leq\pi/4$, with unit amplitude to be multiplied by the lattice codeword. These multipliers will sit in between the integer coefficient $a$ and the signal lattice codeword ${\bf x}$ in the recovered linear %combination of lattice codewords from different transmitters. For example, in the two user case, we have $a_1{\bf x}_1+a_2{\bf x}_2$ which will be replaced by $a_1e^{i\phi_1}{\bf x}_1+a_2e^{i\phi_2}{\bf x}_2$ in a phase precoded %CoF protocol. As a result, the approximant space will be the set of all the vectors with components $ae^{i\phi}$ rather than all the vectors with integer or rational entries. This significantly makes the size of approximants %larger. This fact along with divergent results from Diophantine approximation theory are used to improve DoF and the error performance of CoF protocol.

The {\bf main contributions} of this paper can be summarized in the following two points:
\begin{enumerate}
\item{\em Phase precoding with complex scalar-} We propose phase precoding for CoF protocol to increase the computation rate. We assume that the precoder for each transmitter is a complex scalar $e^{i\phi}$, for some $-\pi/4\leq\phi\leq\pi/4$, multiplying the transmitted lattice codeword. For example, in the two-user case, we send $e^{i\phi_1}{\bf x}_1$ and $e^{i\phi_2}{\bf x}_2$ instead of ${\bf x}_1$ and ${\bf x}_2$. The equivalent channel coefficient vector is ${\bf h}'=(e^{i\phi_1}h_1,~e^{i\phi_2}h_2)$. The precoders should be chosen so that the components of ${\bf h}'$ will be more aligned with Gaussian integers. In other words, ${\bf a}'=(a_1e^{i\phi_1},~a_2e^{i\phi_2})$ is chosen to be aligned with ${\bf h}$. Considering ${\bf a}'$ rather than ${\bf a}$ makes the size of approximants significantly larger. This results in a better alignment of ${\bf h}$ and the new network coefficients providing a higher computation rate, which we call \emph{phase precoded computation rate}. We derive a closed form for the phase precoded computation rate and find the optimum phases to maximize this rate for a given integer vector ${\bf a}$. We also show that the phase precoded computation rate is greater than the computation rate in standard CoF~\cite{Nazer11}. We introduce a new low-complexity algorithm of finding network equation coefficients. In particular, we suggest using of ``Quantized Exhaustive Search (QES)'' algorithm, where our search is based on discretizing the parameter $\alpha$. We note that QES has lower complexity than brute force search proposed in~\cite{Nazer11}.

\item{\em Larger DoF obtained by Diophantine approximation-} We employ the results on Diophantine approximation of~\cite{BBKM02} to prove the existence of phase precoders that achieve high DoF. In particular, we use the divergent part of the Khintchine--Groshev theorem for non-degenerate manifolds~\cite{BBKM02} and simultaneous approximation by algebraic hypersurfaces, to show that the faster convergence speed that we obtain from larger approximant space will help us to increase the DoF of phase precoded CoF protocol. Different from the upper bound on the DoF for CoF in~\cite{Niesen}, we provide a lower bound on the DoF for a phase precoded CoF scheme with $L$ users and $1$ relay node. This lower bound is such that, if the number of transmitters $L$ tends to infinity, then DoF approaches $1$. The other advantage of phase precoded CoF is that no power enhancement occurs.

\end{enumerate}

In Section II, we formulate the problem. In Section III, we propose our phase precoding scheme and analyze its computation rate. The solution to the practical issue of finding network equations is addressed by introducing QES algorithm in this section. In section IV, we investigate the DoF of the proposed phase precoded CoF. Finally, we present concluding remarks in Section V.\\

{\bf Notation.} Boldface letters are used for vectors, and capital boldface letters for matrices. Capital letters are used for functions, and capital calligraphic letters for sets.
Superscripts $^T$ and $^H$ denote transposition and Hermitian transposition. $\mathcal{R}$ denotes a ring and $\mathcal{M}_k(\mathcal{R})$ denotes the set of all $k\times k$ matrices over $\mathcal{R}$. $\mathcal{G}$ and $\mathcal{G}'$ denote a group and its subgroup, and $\mathcal{G}/\mathcal{G}'$ denotes the quotient set. The sets $\mathbb{Z}$, $\mathbb{C}$, $\mathbb{R}$, and $\mathbb{Z}[i]$ denote the ring of rational integers, the field of complex numbers, the field of real numbers, and the ring of Gaussian integers, respectively, where $i^2 = -1$.
We further let $\mathbb{C}^\ast =\mathbb{C}\setminus\{0\}$.

The real and imaginary parts of a complex number are given by $\Re{(\cdot)}$ and $\Im{(\cdot)}$. We let $|z|$ and $\mbox{arg}(z)$ denote the modulus and the unique phase of the complex number $z$, respectively. The Hermitian scalar product of two vectors ${\bf a}$ and ${\bf b}$ is denoted by $\langle{\bf a}, {\bf b}\rangle \triangleq {\bf a} {\bf b}^{H}$. The notations $\|\cdot \|_\infty$ and $\|{\bf v} \|_2$ stand for the maximum norm and the Euclidean norm of a vector ${\bf v}\in\mathbb{R}^n$, respectively. The distance between a vector ${\bf v}\in \mathbb{R}^k$ and a non-empty set $\mathcal{A}\subseteq\mathbb{R}^k$, is defined by $\delta_\infty({\bf v},\mathcal{A})=\inf_{{\bf a}\in\mathcal{A}}\|{\bf v}-{\bf a}\|_\infty,$ where, for an empty set $\mathcal{A}=\phi$, the following relation holds $\delta_\infty({\bf v},\phi)=\infty$. Similarly, we define $\delta_2({\bf v},\mathcal{A})=\inf_{{\bf a}\in\mathcal{A}}\|{\bf v}-{\bf a}\|_2$. Given a positive number $x$, the operator $\log^+(\cdot)$ is defined by $\log^+(x)\triangleq \max\{\log(x),0\}$ For integers $a,b$ and $d$, we denote the relation that $d$ divides $a-b$ by
$a\equiv b\pmod{d}$. Finally, an $k\times k$ matrix ${\bf X}=\left({\bf x}_1^T|\cdots|{\bf x}_k^T \right)^T$ is formed by stacking the $k-$dimensional row vectors ${\bf x}_1,\ldots, {\bf x}_k$, and ${\bf I}_k$ denote the $k\times k$ identity matrix.
%%%%%%%%%%%%%%%%%%%%%%%%%%%%%%%%%%%%%%%%%%%%%%%%%%%%%%%%%%%%%%%%%%%%%%%%%%%%%%%%%%%%%%%%%
\section{Problem Formulation}
%%%%%%%%%%%%%%%%%%%%%%%%%%%%%%%%%%%%%%%%%%%%%%%%%%%%%%%%%%%%%%%%%%%%%%%%%%%%%%%%%%%%%%%%%
We briefly recall the notions of real and complex lattices and corresponding lattice codes, which are essential throughout the paper. A $k$-dimensional {\em complex (real, respectively) lattice} $\Lambda$ with basis $\{{\bf g}_1,{\bf g}_2,\ldots,{\bf g}_k\}$, where ${\bf g}_j \in \mathbb{C}^n$ (${\bf g}_j \in \mathbb{R}^n$, respectively), for $1\leq j\leq k$,
includes points represented as a linear combination of basis vectors with coefficients in $\mathbb{Z}[i]$ ($\mathbb{Z}$, respectively). If $n=k$, the lattice is called {\em full rank}. Let us define the {\em generator matrix} of a full rank lattice $\Lambda$ as the $n\times n$ complex (real, respectively) matrix
$${\bf G}\triangleq \left(\begin{array}{c|c|c|c}
{\bf g}_1^T&{\bf g}_2^T&\cdots &{\bf g}_n^T
\end{array}\right)^T.$$
One can also express $\Lambda$ as $\{{\bf x}={\bf u}{\bf G}| {\bf u}\in \mathbb{Z}[i]^n\}$
($\{{\bf x}={\bf u}{\bf G}| {\bf u}\in \mathbb{Z}^n\}$, respectively).
Around each lattice point is its {\em Voronoi region} consisting of all points of the underlying space which are closer (in terms of complex (real, respectively) Euclidean norm) to that lattice point than other points.
The {\em Gram matrix} of this lattice is ${\bf M}={\bf G}{\bf G}^H$.

A subset $\Lambda'\subseteq\Lambda$ is called a {\em sublattice} if $\Lambda'$ is a lattice itself. Given a sublattice $\Lambda'$, we define the quotient $\Lambda/\Lambda'$ as a {\em lattice code}. This includes a finite constellation of lattice points carved from the lattice $\Lambda$. The shape of such constellation is governed by the Voronoi region of the shaping lattice $\Lambda'$ as explained in \cite{Conway83}. A common choice for the sublattice $\Lambda'$ is $a\Lambda$ for some Gaussian integer $a\in\mathbb{Z}[i]$ ($a\in\mathbb{Z}$, respectively).

For a vector ${\bf y}\in\mathbb{C}^n$, the {\em nearest-neighbor quantizer associated with $\Lambda$} is defined as
\begin{equation}~\label{Q}
Q_{\Lambda}({\bf y}) \triangleq \arg\!\min_{\boldsymbol{\lambda}\in\Lambda}\|{\bf y}-\boldsymbol{\lambda}\|.
\end{equation}
We let $\lfloor{\bf y}\rceil=Q_{\mathbb{Z}}({\bf y})$. We also define the {\em modulo lattice operation} as
$${\bf y}\!\!\!\mod \Lambda\triangleq{\bf y}-Q_{\Lambda}({\bf y}).$$
%%%%%%%%%%%%%%%%%%%%%%%%%%%%%%%%%%%%%%%%%%%%%%%%%%%%%%%%%%%%%%%%%%%%%%%%%%%%%%%%%%%%%%%%%
\subsection{The compute-and-forward protocol}
%%%%%%%%%%%%%%%%%%%%%%%%%%%%%%%%%%%%%%%%%%%%%%%%%%%%%%%%%%%%%%%%%%%%%%%%%%%%%%%%%%%%%%%%%
In \cite{Nazer11} the CoF model is proposed for a relaying network with $L$ transmitters and $M$ relays. The $M$ relays compute estimates of $M$ linear equations of the transmitted information. These are forwarded to the destination, which forms a system of linear equations to recover the distinct messages. It is required that $M\geq L$, in order to be able to solve the system of $M$ linear equations with $L$ unknown variables at the final destination. In the following, for simplicity, we focus on a system model with only one relay node since all relays will operate similarly.

\begin{figure*}[h]
\begin{center}
\includegraphics[width=8cm]{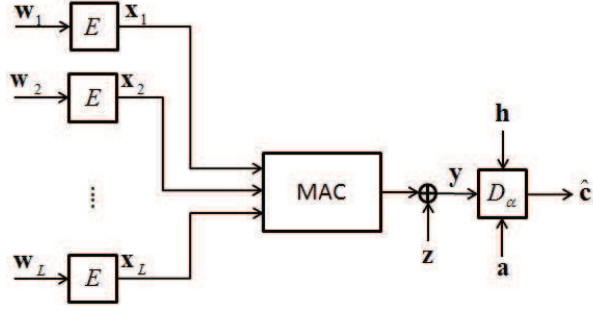}
\caption{\label{CF} $L$ transmitters reliably communicate linear function ${\bf c}=\left(\sum_{\ell=1}^La_\ell{\bf x}_\ell\right)\!\!\!\mod\Lambda'$
to a relay over a complex-valued AWGN network.}
\end{center}
\end{figure*}
Fig.~\ref{CF} illustrates compute-and-forward (CoF) protocol with $L$ transmitters and $1$ relay node. In this scheme, the $\ell$-th transmitter is equipped with an encoder $E:~\mathbb{F}^{k}\rightarrow\Lambda/\Lambda'\subseteq\mathbb{C}^n$, where $k$ is the length of information symbol vector, $\mathbb{F}$ is a finite field of prime size $p$, and $n$ is codeword length. Each encoder maps an information symbol vector ${\bf w}_\ell\in \mathbb{F}^{k}$ to a lattice codeword $E({\bf w}_\ell)={\bf x}_\ell\in\Lambda/\Lambda'$, for $1\leq\ell\leq L$. Each codeword satisfies the power constraint $\|{\bf x}_\ell\|^2\leq n\rho$, where $\rho$ is the power of each symbol. The relay observes a noisy linear combination of the transmitted signals,
\begin{equation}
{\bf y}=\sum_{\ell=1}^Lh_{\ell}{\bf x}_\ell+{\bf z},
\end{equation}
where $h_{\ell}\in\mathbb{C}$, for $1\leq\ell\leq L$, are the Rayleigh fading multiple access channel (MAC) coefficients and ${\bf z}$ is an identically and independently distributed (i.i.d.) Gaussian complex noise $\mathcal{N}_{\mathbb{C}}(0,1)$.

The task of the relay is to estimate a linear combination $\sum_{\ell=1}^La_{\ell}{\bf x}_\ell$ of the transmitted signals given an integer coefficient vector ${\bf a}\triangleq(a_{1},\ldots,a_{L})\in\mathbb{Z}[i]^L$. Due to the linear structure of lattices, the integer linear combinations are still in $\Lambda$ but not necessarily in the lattice code $\Lambda/\Lambda'$. At the relay a detector
\begin{equation}~\label{eq:Decoder}
D_\alpha:\mathbb{C}^L\times\mathbb{C}^n\times\mathbb{Z}[i]^L\rightarrow\Lambda/\Lambda',
\end{equation}
is employed to find an estimate $\hat{\bf c}$ of the codeword linear combination
\[
{\bf c} \triangleq \left(\sum_{\ell=1}^La_{\ell}{\bf x}_\ell\right)\!\!\!\mod{\Lambda'},
\]
which is a point in $\Lambda/\Lambda'$. The quality of this estimation is controlled by a non-zero coefficient $\alpha$. The decoder at the relay first computes
\begin{eqnarray}
\alpha{\bf y} &=& \sum_{\ell=1}^L\alpha h_{\ell}{\bf x}_\ell+\alpha{\bf z}\\
&=&\underbrace{\sum_{\ell=1}^La_{\ell}{\bf x}_\ell}_{\mbox{useful term}}+\underbrace{\sum_{\ell=1}^L\left(\alpha h_{\ell} - a_{\ell}\right){\bf x}_\ell+\alpha{\bf z}}_{\mbox{effective noise}},\label{eq:useful}
\end{eqnarray}
and then sets
$$\hat{\bf c}\triangleq D_{\alpha}({\bf h},{\bf y},{\bf a})= Q_{\Lambda}(\alpha{\bf y}) \!\!\!\mod{\Lambda'},$$
where ${\bf h}\triangleq(h_{1},\ldots,h_{L})\in\mathbb{C}^L$ and $Q_\Lambda$ and $D_\alpha$ are defined in \eqref{Q} and \eqref{eq:Decoder}, respectively. The estimate $\hat{\bf c}$ of ${\bf c}$ will be sent through the network.
In this framework, we declare an {\em equation error} at the relay, if $\hat{\bf c}\neq{\bf c}$. This refers to the event of decoding to an incorrect lattice codeword.

We recall the computation rate for CoF protocol, which is originally defined in~\cite{Nazer11}:
\begin{Proposition}\label{prop:comprate}
For complex-valued AWGN networks with a channel coefficient vector ${\bf h}\triangleq(h_1,\ldots,h_L)\in\mathbb{C}^L$ and a coefficient vector ${\bf a}\in\mathbb{Z}[i]^L$, the following {\em computation rate} is achievable:
\begin{equation}\label{ComputationRate}
\mathfrak{R}(\rho,{\bf h},{\bf a})\triangleq\max_{\alpha\in\mathbb{C}^\ast} \log^+\left(\frac{\rho}{\rho\|\alpha{\bf h}-{\bf a}\|^2+|\alpha|^2}\right).
\end{equation}\hfill$\Box$
\end{Proposition}
From \eqref{eq:useful}, we note that the average energy of the effective noise is
\begin{equation}~\label{eq:q}
Q({\bf a},\alpha)=\rho\|\alpha{\bf h}-{\bf a}\|^2+|\alpha|^2,
\end{equation}
affecting the computation rate. The computation rate, given ${\bf a}$, provided in Proposition~\ref{prop:comprate} is uniquely maximized by choosing $\alpha$ to be the minimum mean square estimator (MMSE) coefficient~\cite{Nazer11}
\begin{equation}\label{alphammse}
\alpha_{\mbox{\tiny MMSE}}=\frac{\rho \langle{\bf h},{\bf a}\rangle}{1+\rho\|{\bf h}\|^2}.
\end{equation}
Substituting $\alpha_{\mbox{\tiny MMSE}}$ of \eqref{alphammse} into $\mathfrak{R}(\rho,{\bf h},{\bf a})$ yields~\cite{Feng}
\begin{equation}~\label{RateGram}
\mathfrak{R}(\rho,{\bf h},{\bf a})=\log^+\left(\frac{1}{{\bf a}{\bf M}{\bf a}^H}\right),
\end{equation}
where ${\bf M}$ is the Gram matrix of a lattice with
\begin{equation}~\label{Gram}
{\bf M}={\bf I}-\frac{\rho}{1+\rho\|{\bf h}\|^2}{\bf h}^H{\bf h}.
\end{equation}
We will address all known solutions to the problem of finding optimum ${\bf a}$ to maximize \eqref{RateGram} in Section~\ref{sec:QES} in more details. We will propose a new low-complexity method of obtaining ${\bf a}$ by discretizing $\alpha$ in Section~\ref{sec:QES}.
%%%%%%%%%%%%%%%%%%%%%%%%%%%%%%%%%%%%%%%%%%%%%%%%%%%%%%%%%%%%%%%%%%%%%%%%%%%%%%%%%%%%%%%%%
\section{Phase Precoding for Compute-and-Forward}
\begin{figure*}[h]
\begin{center}
\includegraphics[width=8cm]{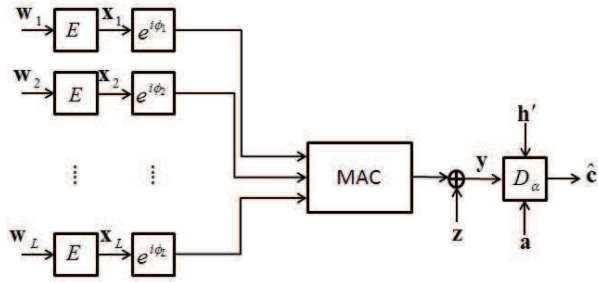}
\caption{\label{CFPP} $L$ transmitters employ different phase precoders and transmit $e^{i \phi_{1}}{\bf x}_1,\ldots,e^{i \phi_{L}}{\bf x}_L$
to a relay over a complex-valued AWGN network.}
\end{center}
\end{figure*}

%%%%%%%%%%%%%%%%%%%%%%%%%%%%%%%%%%%%%%%%%%%%%%%%%%%%%%%%%%%%%%%%%%%%%%%%%%%%%%%%%%%%%%%%%
Fig.~\ref{CFPP} illustrates a network with $L$ transmitters equipped with phase precoders (PP) and $1$ relay employing CoF strategy. After the encoder $E$,
a lattice codeword ${\bf x}_\ell \in\Lambda/\Lambda'$, $1\leq \ell\leq L$, is generated at the $\ell$-th transmitter. We consider a block fading channel model, {\em i.e.} the channel coefficients ${\bf h}$ remain unchanged for a time frame of length $t\gg n$. These channel gains vary independently from one frame to the next. A frame header is used for the training phase, where we apply a phase precoding function $P_\ell\colon\mathbb{C}^n\rightarrow\mathbb{C}^n$, which maps ${\bf x}_\ell$ to $P_\ell({\bf x}_\ell)\triangleq e^{i \phi_{\ell}}{\bf x}_\ell$, for $\phi_\ell\in[-\pi/4,\pi/4]$ and $1\leq \ell\leq L$. Due to the symmetry of the complex plane, the problem of choosing the optimum network equation coefficients for CoF protocol can be reduced to the vectors ${\bf a}$ with components $a_{\ell}$ satisfying $\arg(a_{\ell})\in[-\pi/4,\pi/4]$. Thus, the phases for precoding can also be restricted to $e^{i \phi_{\ell}}$ with $\phi_\ell\in[-\pi/4,\pi/4]$. Using this approach, the phase precoded codeword $e^{i\phi_\ell}{\bf x}_\ell$ continues to satisfy the power constraint $\|e^{i\phi_\ell}{\bf x}_\ell\|^2\leq n\rho$, for $1\leq \ell \leq L$. Thus, the relay receives
\begin{eqnarray}
{\bf y} &=& \sum_{\ell=1}^L h_{\ell}e^{i \phi_{\ell}}{\bf x}_\ell+{\bf z}.\label{eq:Phibetweenhandx}
\end{eqnarray}
We let $h'_{\ell}=h_{\ell}e^{i \phi_{\ell}}$, for $1\leq \ell \leq L$ and
${\bf h}'\triangleq(h'_{1},\ldots,h'_{L})={\bf h}\Phi$,
where
\begin{equation}\label{Phi}
{\Phi}\triangleq\mbox{diag}\left(e^{i \phi_{1}},\ldots,e^{i \phi_{L}}\right).
\end{equation}
As a result of considering the matrix $\Phi$ as part of ${\bf h}'$, the relay recovers an integer linear combination $\sum_{\ell=1}^La_{\ell}{\bf x}_\ell$ of the transmitted codewords. Therefore, it first computes:
\begin{eqnarray}
{\bf y}'&=& \alpha{\bf y} = \sum_{\ell=1}^L\alpha h'_{\ell}{\bf x}_\ell+\alpha{\bf z}\label{eq:PPeffectivenoise}\\
&=& \sum_{\ell=1}^La_{\ell}{\bf x}_\ell+\underbrace{\sum_{\ell=1}^L\left(\alpha h'_{\ell}- a_{\ell}\right){\bf x}_\ell+\alpha{\bf z}.}_{\mbox{PP effective noise}} \nonumber
\end{eqnarray}
The decoder $D_\alpha$ will operate similarly to the CoF protocol except that it assumes ${\bf h}'$ rather than ${\bf h}$. The {\em phase precoded computation rate} for this relay is defined as
\begin{equation}~\label{rateprecoder}
\mathfrak{R}'(\rho,{\bf h},\Phi,{\bf a})=\max_{\alpha\in\mathbb{C}^\ast}\log^+\left(\frac{\rho}{\rho\|\alpha{\bf h}'-{\bf a}\|^2+|\alpha|^2}\right).
\end{equation}
Based on \eqref{eq:PPeffectivenoise}, the average energy of the PP effective noise is
\begin{equation}~\label{firstminimization}
Q'({\Phi},{\bf a},\alpha)=\rho\|\alpha{\bf h}'-{\bf a}\|^2+|\alpha|^2,
\end{equation}
which appears in the denominator of \eqref{rateprecoder}. Therefore, the relay should calculate the best non-zero parameter $\alpha\in\mathbb{C}$ and a non-zero network equation coefficient vector ${\bf a}\in\mathbb{Z}[i]^L$, to maximize \eqref{rateprecoder} or equivalently minimize \eqref{firstminimization}.
%%%%%%%%%%%%%%%%%%%%%%%%%%%%%%%%%%%%%%%%%%%%%%%%%%%%%%%%%%%%%%%%%%%%%%%%%%%%%%%%%%%%%%%%%
\subsection{Maximizing phase precoded computation rate}
%%%%%%%%%%%%%%%%%%%%%%%%%%%%%%%%%%%%%%%%%%%%%%%%%%%%%%%%%%%%%%%%%%%%%%%%%%%%%%%%%%%%%%%%%
To maximize \eqref{rateprecoder}, given ${\bf a}\in\mathbb{Z}[i]^L$,
we first find the optimum $\alpha$ and then the optimum phase precoder ${\Phi}$ using the following lemmas.

\begin{lem}\label{lem:fixedaphi}
Given ${\bf a}\in\mathbb{Z}[i]^L$ and ${\Phi}=\mbox{diag}\left(e^{i \phi_{1}},\ldots,e^{i \phi_{L}}\right)$ with $\phi_\ell\in[-\pi/4,\pi/4]$, and $1\leq\ell\leq L$, the optimum $\alpha_{\tiny\mbox{opt}}'\in\mathbb{C}$ to maximize \eqref{rateprecoder} is
\begin{equation}~\label{alphammseprecoder}
\alpha_{\tiny\mbox{opt}}'=\frac{\rho{\bf a}\Phi^H{\bf h}^H}{1+\rho\|{\bf h}\|^2}=\frac{\rho\langle{\bf h}',{\bf a}\rangle}{1+\rho\|{\bf h}\|^2}.
\end{equation}
\end{lem}
\begin{IEEEproof}
See Appendix~\ref{app:thrate}.
\end{IEEEproof}
Substituting $\alpha_{\tiny\mbox{opt}}'$ into \eqref{rateprecoder} yields:
\begin{eqnarray}%~\label{rateprecoderafter}
\mathfrak{R}'(\rho,{\bf h},{\Phi},{\bf a})&=&\log^+\left(\frac{1}{{\bf a}{\Phi}^H{\bf M}~{\Phi}{\bf a}^H}\right)\label{eq:aphiMphia}\\
&=&\log^+\left(1+\rho\|{\bf h}'\|^2\right)-\log^+\left(\underbrace{\|{\bf a}\|^2+\rho\left(\|{\bf h}'\|^2\|{\bf a}\|^2 -\left|\langle{\bf h}',{\bf a}\rangle\right|^2\right)}_{\mbox{PP Loss Term}}\right)\nonumber,
%&=&\log^+\left(1+\rho\|{\bf h}\|^2\right)-\log^+\left(\|{\bf a}\|^2+\rho\left(\|{\bf h}\|^2\|{\bf a}\|^2 -\left|\langle{\bf h}',{\bf a}\rangle\right|^2\right)\right),\nonumber
\end{eqnarray}
where ${\bf M}$ is given in \eqref{Gram}, ${\bf h}'={\bf h}{\Phi}$, and the second step is based on~\cite{Nazer11,Niesen}.
\begin{lem}\label{lem:fixeda}
Given the network equation coefficients ${\bf a}=(a_1,\ldots,a_L)=(\beta_1e^{i\psi_1},\ldots, \beta_Le^{i\psi_L})\in\mathbb{Z}[i]^L$,
and the channel coefficients ${\bf h}=(h_1,\ldots,h_L)=(\eta_1e^{i\theta_1},\ldots, \eta_Le^{i\theta_L})\in\mathbb{C}^L$,
the optimum phases for $\Phi$ to maximize \eqref{rateprecoder} are $\phi^{\tiny\mbox{opt}}_\ell=\theta_\ell-\psi_\ell$, for $1\leq \ell\leq L$.
In addition, the phase precoded computation rate is greater than the original computation rate, i.e.
$\mathfrak{R}'(\rho,{\bf h},\Phi^{\tiny\mbox{opt}},{\bf a})\geq \mathfrak{R}(\rho,{\bf h},{\bf a})$.
\end{lem}
\begin{IEEEproof}
See Appendix~\ref{app:thrate}.
\end{IEEEproof}
\begin{Theorem}~\label{th:rateprecoder}
Given the channel coefficients ${\bf h}\in\mathbb{C}^L$ and the network equation coefficient vector ${\bf a}\in\mathbb{Z}[i]^L$, the phase precoded computation rate
\begin{equation}\label{eq:final}
\mathfrak{R}'(\rho,{\bf h},\Phi^{\tiny\mbox{opt}},{\bf a})=\log\left(1+\rho\|{\bf h}\|^2\right)-\log\left(\!\|{\bf a}\|^2\!+\!\rho\left(\|{\bf h}\|^2\|{\bf a}\|^2\!-\!\left(\sum_{\ell=1}^L|h_{\ell}||a_{\ell}|\right)^2\right)\right),
\end{equation}
which is greater than $\mathfrak{R}(\rho,{\bf h},{\bf a})$ where
\begin{equation}~\label{eq:phitopopt}
\Phi^{\tiny\mbox{opt}}\triangleq\mbox{diag}\left(e^{i\phi^{\tiny\mbox{opt}}_1},\ldots,e^{i\phi^{\tiny\mbox{opt}}_L}\right).
\end{equation}
\end{Theorem}
\begin{IEEEproof}
See Appendix~\ref{app:thrate}.
\end{IEEEproof}

\noindent
%Theorem \ref{th:rateprecoder} provides the maximal achievable rate at a given signal-to-noise ratio $\rho$. To maximize the phase precoded computation rate, the optimum phase precoder matrix and the corresponding network equation %coefficients should be computed jointly. This is a mixed integer programming problem because the entries of the phase precoding matrix $\Phi$ are complex numbers and the components of ${\bf a}$ are Gaussian integers. In addition, %the phase precoders need to be optimized at the transmitters and the integer coefficients have to be computed at the relay. Recalling Lemma~\ref{lem:fixeda}, for a given ${\bf a}$, the optimum $\Phi^{\tiny\mbox{opt}}$ can be %derived as \eqref{eq:phitopopt}. However, this needs the knowledge of ${\bf a}$ at the transmitters.
%%%%%%%%%%%%%%%%%%%%%%%%%%%%%%%%%%%%%%%%%%%%%%%%%%%%%%%%%%%%%%%%%%%%%%%%%%%%%%%%%%%%%%%%%
\subsection{Quantized exhaustive search (QES)}~\label{sec:QES}
%%%%%%%%%%%%%%%%%%%%%%%%%%%%%%%%%%%%%%%%%%%%%%%%%%%%%%%%%%%%%%%%%%%%%%%%%%%%%%%%%%%%%%%%%
Using \eqref{eq:aphiMphia} for a fixed $\Phi$, a method of finding the optimum ${\bf a}$ is to consider ${\bf M}' = \Phi^H{\bf M}\Phi$ and employ one of the approaches presented in~\cite{Feng,sakzad12,Nazer11,caire}. This means that the optimum ${\bf a}$ can only be computed at the relays when the optimum $\Phi$ was known at the transmitters. Next we review the available approaches of finding the optimum ${\bf a}$ for a fixed $\Phi$ and then introduce a new algorithm by discretizing $\alpha$. We also compare the computational complexity of our proposed algorithm in comparison with other methods.

Recall from \eqref{rateprecoder} and \eqref{eq:aphiMphia} that
\begin{eqnarray}
\mathfrak{R}'(\rho,{\bf h},{\Phi},{\bf a})&=&\max_{\alpha\in\mathbb{C}^\ast}\log^+\left(\frac{\rho}{\rho\|\alpha{\bf h}'-{\bf a}\|^2+|\alpha|^2}\right)=\max_{\alpha\in\mathbb{C}^\ast}\log^+\left(\frac{\rho}{Q'(\Phi,{\bf a},\alpha)}\right)\nonumber\\
&=&\log^+\left(\frac{1}{{\bf a}{\Phi}^H{\bf M}~{\Phi}{\bf a}^H}\right)\label{eq:replacedalphastar}\\
&=& \log^+\left(\frac{1}{{\bf a}{\bf M}'{\bf a}^H}\right),\nonumber
\end{eqnarray}
where ${\bf M}'=\Phi^H{\bf M}\Phi$ is the Gram matrix of a new lattice $\Lambda'$ totally different from the lattice $\Lambda$ with Gram matrix ${\bf M}$. Note that to obtain \eqref{eq:replacedalphastar} we replaced $\alpha$ by $\alpha_{\tiny\mbox{opt}}'$ from \eqref{alphammseprecoder}. Thus,
for a given $\Phi$ and in order to maximize $\mathfrak{R}'(\rho,{\bf h},{\Phi},{\bf a})$, one should solve the following optimization problem:
\begin{equation}~\label{eq:minprob}
\min_{{\bf a}\in\mathbb{Z}[i]^L}{\bf a}{\bf M}'{\bf a}^H.
\end{equation}
For $\Phi={\bf I}_L$ ({\em i.e.} ${\bf M}'={\bf M}$), the problem of finding the optimum ${\bf a}$ is addressed in~\cite{Feng, caire, Nazer11, Sakzad13-1, sakzad12}.
We summarize these approaches below:
\begin{enumerate}
\item Note that for a given vector ${\bf h}$, the computation rate $\mathfrak{R}'(\rho,{\bf h},{\bf I}_L,{\bf a})=\mathfrak{R}(\rho,{\bf h},{\bf a})$ is {\em zero} if the coefficient vector ${\bf a}$ satisfies~\cite{Nazer11} $$\|{\bf a}\|^2\geq 1+\rho\|{\bf h}\|^2.$$
    The authors of~\cite{Nazer11} have suggested an exhaustive search within a sphere of radius $1+\rho\|{\bf h}\|^2$ to find the optimum vector ${\bf a}$. The optimum $\alpha$ is $\alpha_{\mbox{\tiny MMSE}}$ of \eqref{alphammse}. However this approach is not practical due to its exponential complexity depending on the signal-to-noise ratio $\rho$.
\item In~\cite{Feng}, based on \eqref{eq:minprob}, the optimal value of ${\bf a}$ can be found as
    \begin{eqnarray}
    {\bf a}_{\tiny\mbox{opt}}&=&\arg\!\!\!\!\!\!\!\min_{{\bf 0}\neq{\bf a}\in\mathbb{Z}[i]^L}~{\bf a}{\bf M}{\bf a}^H\label{a_opt}\\
    &=&\arg\!\!\!\!\!\!\!\min_{{\bf 0}\neq{\bf a}\in\mathbb{Z}[i]^L}~\|{\bf a}{\bf L}\|^2\nonumber,
    \end{eqnarray}
    where ${\bf L}$ is the Cholesky decomposition of ${\bf M}'={\bf M}={\bf L}{\bf L}^H$. To find the best ${\bf a}$, a sphere decoder~\cite{ViB} with initial radius $1+\rho\|{\bf h}\|^2$ can be used~\cite{Feng}.
\item In~\cite{sakzad12}, we proposed the use of an $(L+1)$-dimensional lattice $\Lambda_0$ to find appropriate coefficient vector ${\bf a}$. In particular, a complex version of LLL algorithm has been applied to the generator matrix of $\Lambda_0$ and ${\bf a}$ is chosen to be the coefficient vector corresponding to the shortest vector of the new lattice. The corresponding $\alpha$ is found from \eqref{alphammse}. The effectiveness of this method along with two other exhaustive search based algorithms are also studied.
\item A new approach is also introduced in~\cite{caire}. In this framework, LLL algorithm has been employed to reduce the generator matrix ${\bf L}$ and obtain ${\bf L}'$. Then the length of the shortest row of ${\bf L}'$ (in terms of Euclidean norm) is used as the initial radius in Fincke-Pohst algorithm~\cite{Fincke} rather than $1+\rho\|{\bf h}\|^2$~\cite{Nazer11} to seacrh for the optimum ${\bf a}$.
\end{enumerate}
We now present a generalized method of~\cite{sakzad12} to find the best ${\bf a}'_{\tiny\mbox{opt}}$ and $\alpha'_{\tiny\mbox{opt}}$ simultaneously with a reduced complexity. We explain our method for ${\bf M}'$. To recreate the approach for ${\bf M}$, one only need to set $\Phi={\bf I}_L$.

We let $|\alpha|$ vary from $1$ to a maximum value $|\alpha|_{\max}$ in integer steps and the phase of $\alpha$ vary from $0$ to $90$ degrees with a step size of $d$ degrees (e.g., $d=5$), since the ring of Gaussian integers is invariant by a rotation of 90 degrees. To obtain the best network equation coefficients ${\bf a}$, we apply the following quantization $a_{\ell}=Q_{\mathbb{Z}[i]}\left(\alpha h_{\ell}e^{i\phi_\ell}\right)$, for $1\leq\ell\leq L$. For the current vector ${\bf a}$, we then compute $\alpha$ based on \eqref{alphammseprecoder}. Then we select the optimum ${\bf a}'_{\tiny\mbox{opt}}$ and $\alpha'_{\tiny\mbox{opt}}$, which minimizes $\rho\|\alpha{\bf h}'-{\bf a}\|^2+|\alpha|^2$. Note that in this case ${\bf a}'_{\tiny\mbox{opt}}$ and $\alpha'_{\tiny\mbox{opt}}$ are chosen concurrently. This method results in the lowest quantization error plus noise $Q'(\Phi,{\bf a},\alpha)$
after generating a large number of possible ${\bf a}$'s produced the discretized set of $\alpha$'s. Note that our phase precoding scheme as well as the QES algorithm can also be implemented on CoF over Eisenstein integers $\mathbb{Z}[\omega]$,~\cite{SunYuan12,Tunali12}.
A QES routine is given below.

\begin{algorithmic}[1]
\Function{QES}{$\Phi,{\bf h},\rho$}
\State ${\bf h}' \gets {\bf h}\Phi$
\State $q \gets \mbox{Inf}$
\For{$|\alpha|\gets1:1:|\alpha|_{\max}$}
   \For{$\arg(\alpha)\gets 0:d:90$ }
       \State $\alpha\gets|\alpha|e^{i\arg(\alpha)}$
       \State ${\bf a}\gets Q_{\mathbb{Z}[i]^L}\left(\alpha {\bf h}' \right)$
       \State $\alpha \gets \frac{\rho{\bf a}\Phi^H{\bf h}^H}{1+\rho\|{\bf h}\|^2}$
       \State $q_1\gets\rho\|\alpha{\bf h}'-{\bf a}\|^2+|\alpha|^2$
       \If{$q_1 < q$ and ${\bf a}\neq{\bf 0}$}
           \State ${\bf a}'_{\tiny\mbox{opt}}\gets{\bf a}$
           \State ${\alpha}'_{\tiny\mbox{opt}} \gets\alpha$
           \State $ q \gets q_1$
       \EndIf
   \EndFor
\EndFor
\State \textbf{return} ${\bf a}'_{\tiny\mbox{opt}}\in\mathbb{Z}[i]^L$ and $\alpha'_{\tiny\mbox{opt}}\in\mathbb{C}$
\EndFunction
\end{algorithmic}
We note that QES is slightly different from the ``simple quantized search'' presented in~\cite{sakzad12}. The difference is in the evaluation step of MMSE relation, see line $7$ of the function QES. In~\cite{sakzad12}, there is no updating of $\alpha$ before computing the phase precoding effective noise $Q'(\Phi,{\bf a},\alpha)=\rho\|\alpha{\bf h}'-{\bf a}\|^2+|\alpha|^2$. However, we added this step in QES.

%%%%%%%%%%%%%%%%%%%%%%%%%%%%%%%%%%%%%%%%%%%%%%%%%%%%%%%%%%%%%%%%%%%%%%%%%%%%%%%%%%%%%%%%%
\subsection{Complexity comparison}
%%%%%%%%%%%%%%%%%%%%%%%%%%%%%%%%%%%%%%%%%%%%%%%%%%%%%%%%%%%%%%%%%%%%%%%%%%%%%%%%%%%%%%%%%
We compare the complexity of the proposed approach and the exhaustive search algorithm~\cite{Nazer11}.
\begin{enumerate}
\item
In~\cite{Nazer11}, the complexity of the brute force search is of order $O\left(\rho^n\right)$. This is because the search is made over all possible non-zero integer vectors ${\bf a}$, with square norms less than $1+\rho\|{\bf h}\|^2$.
This search approach has the highest complexity.
\item
The search space for $|\alpha|$ is upper bounded by $90/d\times |\alpha|_{\max}$. Hence the complexity order of QES is  $O(90/d\times |\alpha|_{\max})$.
\item
The lattice-reduction-based algorithms presented in \cite{Feng} and \cite{sakzad12} for finding integer network coefficients have polynomial complexity in terms of $L$, which is much lower than the above two approaches. However, obtaining the optimum ${\bf a}'_{\tiny\mbox{opt}}$ is not guaranteed for system models with $L>2$ as both the real and complex LLL algorithms give the exact solution to \eqref{eq:minprob} only for $L=2$.
\end{enumerate}

%%%%%%%%%%%%%%%%%%%%%%%%%%%%%%%%%%%%%%%%%%%%%%%%%%%%%%%%%%%%%%%%%%%%%%%%%%%%%%%%%%%%%%%%%
\section{Degrees of Freedom for Phase Precoded CoF protocol}
%%%%%%%%%%%%%%%%%%%%%%%%%%%%%%%%%%%%%%%%%%%%%%%%%%%%%%%%%%%%%%%%%%%%%%%%%%%%%%%%%%%%%%%%%
Let us define $\mathfrak{R}(\rho,{\bf h})\triangleq\sup_{\bf a}\mathfrak{R}(\rho,{\bf h},{\bf a})$,
then the {\em degrees-of-freedom of the CoF}  rate is defined as~\cite{Niesen}:
$$\limsup_{\rho\rightarrow\infty}\frac{\mathfrak{R}(\rho,{\bf h})}{\log (\rho)}.$$
The {\em degrees-of-freedom of the phase precoded CoF} can be defined similarly as
$$\limsup_{\rho\rightarrow\infty}\frac{\mathfrak{R}'(\rho,{\bf h})}{\log (\rho)},$$
where
$\mathfrak{R}'(\rho,{\bf h})\triangleq\sup_{\Phi}\sup_{\bf a}\mathfrak{R}'(\rho,{\bf h},\Phi,{\bf a})$.

Niesen {\em et al.}~\cite{Niesen} showed that one can restate \eqref{RateGram} as follows:
\begin{eqnarray}
\mathfrak{R}(\rho,{\bf h},{\bf a})&=&\frac{1}{2}\log\left(1+\rho\|{\bf h}\|^2\right)\nonumber\\
&-&\frac{1}{2}\log\left(\underbrace{\|{\bf a}\|^2+\rho\left(\|{\bf h}\|^2\|{\bf a}\|^2-\left|\langle{\bf h},{\bf a}\rangle\right|^2\right)}_{\mbox{Loss Term}}\right),\label{lossterm}
\end{eqnarray}
where the first term is the capacity of a MAC with channel coefficients ${\bf h}$ and the second term is the penalty of approximating the channel gains by integer numbers. This cost equals to $\frac{1}{2}\log\left(\|{\bf a}\|^2\right)$ if and only if ${\bf a}$ is a multiple scalar of ${\bf h}$. However, this scenario may not occur since ${\bf a}$ has integer components and the entries of ${\bf h}$ are complex numbers. This result shows that the CoF scheme has low DoF. We recall from~\cite{Niesen} that:
\begin{Proposition}
For  almost every ${\bf h}\in\mathbb{R}^L$ ({\em i.e.} the set of all ${\bf h}\in\mathbb{R}^L$ that do not satisfy the following inequality are of Lebesgue measure zero), we have
\begin{equation}~\label{eq:DoFO}
\limsup_{\rho\rightarrow\infty}\frac{\max_{{\bf 0}\neq{\bf a}\in\mathbb{Z}^L}\mathfrak{R}(\rho,{\bf h},{\bf a})}{\frac{1}{2}\log (\rho)}\leq \left\{ \begin{array}{cl}  \frac{1}{2}  & L=2,\rule[-4mm]{0mm}{7mm}\\   \frac{2}{L+1} & L>2.\end{array} \right.
\end{equation}\hfill$\Box$
\end{Proposition}
Khintchine's theorem from the theory of Diophantine approximation plays a central role in the proof of the above result. The authors of~\cite{Niesen} then proposed {\em signal alignment} for an $L\times L$ section of a network and used a generalization of the convergence part of Khintchine theorem, called Khintchine--Groshev theorem for non-degenerate manifolds~\cite{B01}, and proved that the full DoF can be obtained in the presence of CSIT. In the following, we investigate the DoF of phase precoded CoF. We will use the divergent part of the Khintchine--Groshev theorem for non-degenerate manifolds~\cite{BBKM02} and the simultaneous approximation by algebraic hypersurfaces for analysis purposes.

We consider the standard conversion method  to transform an $L \times L$ complex matrix ${\bf M}$ given in \eqref{Gram}
to its corresponding $2L \times 2L$ real version:
\begin{equation}~\label{eq:cplx_to_real}
\tilde{\bf M}\triangleq\left(
\begin{array}{cc}
\Re\left({\bf M}\right)& \Im\left({\bf M}\right)\\
-\Im\left({\bf M}\right)& \Re\left({\bf M}\right)
\end{array}\right).
\end{equation}
Similarly, a complex vector ${\bf a}$ (respectively, ${\bf h}$) will be replaced by $(\Re\left({\bf a}\right), \Im\left({\bf a}\right) )$.
For simplicity we will continue to denote such a vector as ${\bf a}$.
As a result, the phase precoded computation rate \eqref{eq:aphiMphia} can be rewritten as
\begin{equation}~\label{eq:newmin1}
\mathfrak{R}'(\rho,{\bf h},\tilde{\Phi},{\bf a})= \log^+\left(\frac{1}{{\bf a}\tilde{\Phi}^T\tilde{\bf M}\tilde{\Phi}{\bf a}^T}\right),
\end{equation}
where the rotation matrix $\tilde{\Phi}$ obtained by converting the matrix
$\Phi=\mbox{diag}\left(e^{i \phi_{1}},\ldots,e^{i \phi_{L}}\right)$ to its real version as in \eqref{eq:cplx_to_real}.
To maximize \eqref{eq:newmin1}, we need to solve the optimization problem:
\begin{equation}~\label{eq:newmin}
\min_{\tilde{\Phi}}\min_{{\bf 0}\neq{\bf a}\in\mathbb{Z}^{2L}}{\bf a}\tilde{\Phi}^T\tilde{\bf M}\tilde{\Phi}{\bf a}^T.
\end{equation}
We let ${\bf a}'={\bf a}\tilde{\Phi}^T$ and solve
\begin{equation}~\label{eq:newminanalysis}
\min_{\tilde{\Phi}}\min_{{\bf 0}\neq{\bf a}'} {\bf a}'\tilde{\bf M}\left({\bf a}'\right)^T~.
\end{equation}

We now show the existence of a phase precoder which guarantees a lower bound on the DoF of CoF.
\begin{Theorem}~\label{th:DoF}
For almost every channel coefficients ${\bf h}\in\mathbb{R}^{2L}$, there exists a phase precoder matrix $\tilde{\Phi}$ that gives
\begin{equation}~\label{DoF}
\limsup_{\rho\rightarrow\infty}\frac{\max_{{\bf a}\in\mathbb{Z}^{2L}}\mathfrak{R}'(\rho,{\bf h},\tilde{\Phi},{\bf a})}{\frac{1}{2}\log \rho}\geq\frac{2L+2}{2L+2+\frac{2}{c_2}},
\end{equation}
where $c_2$ is a positive constant independent of $L$. Furthermore, if $L\rightarrow\infty$, then the achievable DoF for a phase precoded CoF protocol tends to be one.
\end{Theorem}
\begin{IEEEproof}
See Appendix~\ref{app:thDoF}.
\end{IEEEproof}
Note that the above theorem provides a lower bound on the DoF of phase precoded CoF whereas the previous results \cite{Niesen} only provided upper bounds. In particular, comparing \eqref{DoF} to \cite[eq.~(21)]{Niesen}, we also observe that DoF of phase precoded CoF obtained from our lower bound is much greater than the DoF in the upper bound with no precoding, as $L$ increases.

%%%%%%%%%%%%%%%%%%%%%%%%%%%%%%%%%%%%%%%%%%%%%%%%%%%%%%%%%%%%%%%%%%%%%%%%%%%%%%%%%%%%%%%%%%
\section{Conclusion and Further Research Topics}
%%%%%%%%%%%%%%%%%%%%%%%%%%%%%%%%%%%%%%%%%%%%%%%%%%%%%%%%%%%%%%%%%%%%%%%%%%%%%%%%%%%%%%%%%
A phase precoder scheme has been introduced for compute-and-forward (CoF) protocol in physical layer network coding~\cite{Liew11}. The properties of this scheme has been investigated in terms of achievable rate, degrees-of-freedom (DoF), and error rate performance. It has been proven that there exists a phase precoder such that DoF approaches $1$ for almost every channel realization. We employed Diophantine approximation to prove that our larger approximant space associated to phase precoding increases the DoF of CoF. To simplify the implementation, we proposed a new approach, referred to as QES algorithm, to find the network equation coefficients. The other advantage of using phase precoding in CoF is that no power enhancement occurs.
%We introduced a practical phase precoder with limited feedback based on deep holes of Gaussian integer lattice. To simplify the implementation, we proposed a new approach, referred to as QES algorithm, to find the network equation coefficients. To validate the efficiency of our scheme, computer simulation results were presented. For example, a coding gain of $2.5$dB at EER of $10^{-5}$ has been obtained using phase precoding over an $E_8/4E_8$ Voronoi lattice encoder.

The idea of phase precoding for CoF has also been extended to integer-forcing MIMO linear receivers, where unitary precoders are employed to achieve full-diversity~\cite{sakzad14-2}. The phase precoded CoF along with $E_8$ lattice has also been investigated in terms of probability of equation error in~\cite{sakzad14-1}. Using efficient and strong high-dimensional lattice codes carved from the well-known LDPC lattices, LDLC lattices, turbo lattices and LDA lattices~\cite{LDA,LDPCLattice,TL,LDLC} rather than $E_8$ is a promising research problem. Designing lattices and lattice codes matched to CoF and phase precoded CoF is another future research direction.

In a network scenario, at the destination side, the end user must solve a system of linear equations over a finite field $\mathbb{F}_p$. The probability of having a singular matrix over this field is proportional to the inverse of $p$ which results in an error floor in the probability of error. Hence, designing network equations such that they form an invertible matrix at the final destination is of interest for further research~\cite{viterbo11}.
%%%%%%%%%%%%%%%%%%%%%%%%%%%%%%%%%%%%%%%%%%%%%%%%%%%%%%%%%%%%%%%%%%%%%%%%%%%%%%%%%%%%%%%%%
\appendices
%%%%%%%%%%%%%%%%%%%%%%%%%%%%%%%%%%%%%%%%%%%%%%%%%%%%%%%%%%%%%%%%%%%%%%%%%%%%%%%%%%%%%%%%%
%%%%%%%%%%%%%%%%%%%%%%%%%%%%%%%%%%%%%%%%%%%%%%%%%%%%%%%%%%%%%%%%%%%%%%%%%%%%%%%%%%%%%%%%%
\section{Background on Simultaneous Diophantine Approximation by Algebraic Hypersurfaces}~\label{backDio}
%%%%%%%%%%%%%%%%%%%%%%%%%%%%%%%%%%%%%%%%%%%%%%%%%%%%%%%%%%%%%%%%%%%%%%%%%%%%%%%%%%%%%%%%%
\begin{Definition}
A number $\beta\in\mathbb{R}$ is called an algebraic number if $F(\beta)=0$ for a polynomial $F(x)\in\mathbb{Q}[x]$. It is called an algebraic integer if $F$ is a monic polynomial in $\mathbb{Z}[x]$. Let $d$ be the least degree of a polynomial such that $F(\beta)=0$, then $\beta$ is an algebraic number of degree $d$.
\end{Definition}
\begin{Example}
Rational numbers $p/q$ are all algebraic numbers of degree $1$ because $F(p/q)=0$ for $F(x)=qx-p$; The irrational number $\sqrt{2}$ is an algebraic integer of degree $2$ because $G(\sqrt{2})=0$ for $G(x)=x^2-2$.
%The set of algebraic numbers form a ring, {\em i.e.} the product and addition of every two algebraic number is another algebraic number. Furthermore, let $F(\beta)=0$
%for a polynomial $F\in\mathbb{R}[x]$ where all of its coefficients are algebraic numbers, then $\beta$ is also algebraic number.
\end{Example}
\begin{Definition}
For any ${\bf x}\in\mathbb{R}^k$ and positive $\varepsilon$, the $k$-dimensional ball $\mathcal{B}_k({\bf x},\varepsilon)$ is the set of all points ${\bf y}\in\mathbb{R}^k$ such that $\|{\bf x}-{\bf y}\|^2\leq\varepsilon$. We drop $k$ in case of no ambiguity.
\end{Definition}
\begin{Definition}~\label{openset}
A set $\mathcal{O}\subseteq\mathbb{R}^k$ is called open if for every ${\bf x}\in\mathcal{O}$ there exists $\varepsilon>0$ such that $\mathcal{B}({\bf x},\varepsilon)\subseteq\mathcal{O}$.
\end{Definition}
\begin{Definition}~\label{opencover}
A family of open sets $\{\mathcal{O}_\beta\subseteq \mathbb{R}^k\}_{\beta\in\mathcal{I}}$ indexed by the {\em index set} $\mathcal{I}$, is called an open cover of a subset $\mathcal{K}\subseteq \mathbb{R}^k$,
if
$$\mathcal{K}\subseteq \bigcup_{\beta\in\mathcal{I}}\mathcal{O}_\beta.$$
\end{Definition}
\begin{Definition}~\label{compactset}
If every open cover of $\mathcal{K}\subseteq\mathbb{R}^k$ has a finite subcover, then $\mathcal{K}$ is called a compact set.
\end{Definition}
\begin{Example}
All open intervals $(a,b)\subseteq\mathbb{R}$ are open. Any finite union of disjoint closed intervals $\{[a_\beta,b_\beta]\colon a_\beta,b_\beta\in\mathbb{R}\}_{\beta\in\mathcal{I}}$ with $|\mathcal{I}|<\infty$ is a compact set.
It is clear from the above definitions that a single point $\{{\bf v}\}\subseteq\mathbb{R}^k$ is a compact set. In fact, single points are compact in every topology.
\end{Example}
In this section we deal with polynomials and functions of several variables $\{x_1,\ldots,x_k\}$. We use multiindex notation ${\bf x}^{\bf e}$ for $x_1^{e_1}\cdots x_k^{e_k}$ where ${\bf e}=(e_1\ldots,e_k)\in\mathbb{N}^k$.
\begin{Definition}~\label{AlgebraicHypersurfaces}
If $F({\bf x}) \in\mathbb{Z}[{\bf x}]$ is a non-zero polynomial, then $\mathcal{A}(F)$ denotes an algebraic hypersurface consisting of points ${\bf a}\in\mathbb{R}^{k}$ such that $F({\bf a})=0$.
\end{Definition}
\begin{Definition}
Let $\mathcal{M}$ be a finite, nonempty set of functions, then $\mathcal{P}(\mathcal{M})$ denotes the set of nonzero functions $F({\bf x}) \in\mathbb{Z}[{\bf x}]$, which are all integer linear combinations of the functions in $\mathcal{M}$.
\end{Definition}
\begin{Example}~\label{ex:Mmultivariatepoly}
Let $\mathcal{M}$ include all monomials of the type ${\bf x}^{\bf e}=x_1^{e_1}\cdots x_k^{e_k}$. Then $\mathcal{P}(\mathcal{M})$ is the set of nonzero polynomials $F({\bf x}) \in\mathbb{Z}[{\bf x}]$, which are linear combinations of the monomials of $\mathcal{M}$. In the special case where
$\mathcal{M}=\left\{1,x,x^2,\ldots,x^{m-1}\right\}$, then $\mathcal{P}(\mathcal{M})$ is the set of all polynomials with integer coefficients of degree up to $m-1$. For a fixed $F\in\mathcal{P}(\mathcal{M})$, $\mathcal{A}(F)$ is the set of all algebraic numbers $\zeta$ with $F(\zeta)=0$.
\end{Example}
A more general concept of an algebraic hypersurface can also be defined over $\mathcal{P}(\mathcal{M})$ as follows:
\begin{Definition}~\label{GeneralAlgebraicHypersurfaces}
Let $\mathcal{M}$ be a finite, nonempty set of $k$-variate functions, then
$$\mathcal{A}(\mathcal{M})=\{{\bf a}\in\mathbb{R}^k\colon F({\bf a})=0,~\forall F\in\mathcal{P}(\mathcal{M})\}.$$
\end{Definition}
We later consider a broader set $\mathcal{A}_k(\Phi,\mathcal{M})$ (see \eqref{AkphiM}), for which $F({\bf a})$ is approximately zero for all $F\in\mathcal{P}(\mathcal{M})$.
\begin{Definition}~\label{linind}
The set $\mathcal{M}$ with cardinality $m$ of $k$-variate functions $\{G_1,\ldots,G_m\}$ over several variables $x_1,\ldots,x_k$ is called linearly independent if $c_1G_1({\bf x})+\cdots+c_mG_m({\bf x})=0$ for all ${\bf x}=(x_1,\ldots,x_k)$ for some constants $c_1,\ldots,c_m\in\mathbb{R}$, then $c_1=\cdots=c_m=0$.
\end{Definition}
\begin{Definition}~\label{AnalyticFunctions}
Let $\mathcal{T}$ be a subset of $\mathbb{R}^k$. A multivariate function $G\colon \mathcal{T}\rightarrow\mathbb{R}^t$,
with the following mapping formula
$$(x_1,\ldots,x_k)\mapsto\left(G_1(x_1,\ldots,x_k),\ldots,G_t(x_1,\ldots,x_k)\right),$$ is called an analytic function if for any ${\bf x}\in\mathcal{T}$ there exist a ball $\mathcal{B}({\bf x},r)$ such that one can write
$$G({\bf y})=\sum_{\bf e}c_{\bf e}\left({\bf y}-{\bf x}\right)^{\bf e},$$
for every ${\bf y}\in\mathcal{B}({\bf x},r)\cap\mathcal{T}$. Note that this corresponds to the multivariate Taylor expansion of $G$ around ${\bf x}\in\mathcal{T}$.
\end{Definition}
\begin{Example}~\label{ex:Mlinindanalytics}
Let $k=2L$, $m=L+1$, and
$$\mathcal{T}=\{(x_\ell,x_{\ell+L})\in\mathbb{R}^2\colon x_\ell^2+x_{\ell+L}^2=r,~\mbox{where $r$ is the sum of two integer squares}\}.$$
Let us also suppose that $G_\ell:\mathcal{T}\rightarrow\mathbb{R}^2$ maps $(x_\ell,x_{\ell+L})$ to $x_\ell^2+x_{\ell+L}^2$ and
$$\mathcal{M}=\left\{G_\ell\colon~\ell=1,\ldots,L\right\}\cup\{1\}.$$
It is clear that $|\mathcal{M}|=L+1$ and all the polynomials in $\mathcal{M}$ are analytic because they are multivariate polynomials and the third order derivative of every $G_\ell$ is zero. The set $\mathcal{M}$ is trivially a set of linearly independent functions because the elements of $\mathcal{M}$ have different variables. Then $\mathcal{P}(\mathcal{M})$ is the set of nonzero polynomials $F({\bf x}) \in\mathbb{Z}[{\bf x}]$ of the form $$c_1\left(x_1^2+x_{L+1}^2\right)+\cdots+c_{L}\left(x_{L}^2+x_{2L}^2\right)+c_{L+1},$$
for $c_1,\ldots,c_{L+1}\in\mathbb{Z}$.
\end{Example}
\begin{Definition}
For a polynomial
$$F({\bf x})=\sum c_{\bf e} {\bf x}^{\bf e},$$
we define
the {\em Height} of $F$ as
\begin{equation}~\label{hiegth}
H(F) = \max_{\bf e} |c_{\bf e}|.
%H(F) = \max_{1\leq \ell \leq L+1}\{c_\ell\}.
\end{equation}
\end{Definition}
For comments, open problems, and application of Height of a polynomial we refer the reader to~\cite{Borwein02}. The following theorem describes the behavior of multivariate function $F$ around its zeros.
\begin{Theorem}[Lojasiewicz's inequality,~\cite{Lojasiewicz}]\label{th:Lojasiewicz}
For any compact set $\mathcal{K}\subseteq\mathbb{R}^n$ and multivariate function $F$, there are positive constants $c_1$ and $c_2$ such that
\begin{equation}~\label{eq:Lojasiewicz}
\delta_2({\bf v},\mathcal{A}(F))^{c_2}\leq c_1 |F({\bf v})|
\end{equation}
for every ${\bf v}\in\mathcal{K}$.
\end{Theorem}
Note also that the constants $c_1$ and $c_2$ depend on $F$ and can be large. Later we will use the above theorem and the Height of a polynomial to upper bound a quantity in our proof.

The history of Diophantine approximation is quite interesting where it ranges from the estimate of $\pi$ to the theory of continued fractions~\cite{Waldshmidt11}. The theory of Diophantine approximation deals with the approximation of real vectors by a specific subset of real vectors. A familiar related concept in engineering is quantization, where real numbers are approximated with finite precision numbers (quantized values). In our problem the quantized values will be replaced by more general algebraic hypersurfaces and their zeros. In Diophantine approximation theory, there are two types of results: asymptotic and uniform. The ``asymptotic'' results only deal with the number of solutions (which is typically infinitely many) to a specific inequality. These results are valid for all (irrational) numbers and vectors~\cite{Waldshmidt11}. For example, we have
\begin{Theorem}[Asymptotic Dirichlet's Theorem,~\cite{Waldshmidt11}]~\label{th:ADT}
Let $\xi\in\mathbb{R}$ be irrational, then there exist infinitely many $\frac{p}{q}\in\mathbb{Q}$ such that
$$\left|\xi-\frac{p}{q}\right|<\frac{1}{q^2}.$$
\end{Theorem}
Note that for a finite fixed $q$, as in practical quantization, there will be only finitely many rational numbers $\frac{p}{q}$ for which $\left|\xi-\frac{p}{q}\right|<\frac{1}{q^2}$ holds. Other results of Diophantine approximation theory deal with ``almost'' all numbers in the sense that the subset of numbers in $\mathbb{R}$ that can be approximated, with a given decaying error, has either full or null Lebesgue measure. The Diophantine approximation theory that works with almost all numbers is called metric number theory. This theory usually deals with results which have two parts depending on a given metric function ${\it\Psi}$ which defines the decaying behavior of the approximation error. The divergence or convergence of a series with respect to the metric function ${\it\Psi}$ discriminates if the approximation error to almost all numbers decays according to the metric function or not~\cite{Scmidtbook}.
\begin{Theorem}[Uniform Khintchine's Theorem,~\cite{Waldshmidt11}]\label{th:UKT}
Given a non-increasing metric function ${\it\Psi}:\mathbb{N}\rightarrow(0,+\infty)$, let
$$\mathcal{A}({\it\Psi})=\left\{\xi\in[0,1]\colon~\left|q\xi-p\right|<{\it \Psi}\left(q\right)~\mbox{for~infinitely~many}~\frac{p}{q}\in\mathbb{Q}\right\}.$$
Then
\begin{equation}%~\label{eq:metricalresult1dim}
|\mathcal{A}({\it\Psi})|=
\left\{\begin{array}{ll}
0,&\mbox{if}~\sum_{h=1}^\infty{\it\Psi}(h)<\infty,\\
1,&\mbox{if}~\sum_{h=1}^\infty{\it\Psi}(h)=\infty,\\
\end{array}
\right.
\end{equation}
where $|\mathcal{A}({\it\Psi})|$ denotes the Lebesgue measure of $\mathcal{A}({\it\Psi})$.
\end{Theorem}
In other words, if the series $\sum_{h=1}^\infty{\it\Psi}(h)$ is divergent, then for almost all numbers $\xi\in[0,1]$, there are infinitely many rationals $\frac{p}{q}\in\mathbb{Q}$ such that $\left|q\xi-p\right|<{\it \Psi}\left(q\right)$. Another way of thinking about Theorem~\ref{th:UKT} is that the metric function ${\it\Psi}$ gives the accuracy of the approximation as a function of the denominator of the approximant $\frac{p}{q}$. The speed of ${\it\Psi}$ requiring too rapidly decreasing approximation error can lead to the fact that almost all numbers can not be approximated infinitely many times.

We now provide the generalized versions of the above results for the case, where we approximate our real vectors with algebraic hypersurfaces rather than discrete points. Given a metric function ${\it\Psi}:\mathbb{N}\rightarrow(0,+\infty)$, let
\begin{multline}\label{AkphiM}
\mathcal{A}_k({\it\Psi,\mathcal{M}})=\left\{{\bf a}\in[0,1]^k\colon|F({\bf a})|<H(F)^{-m+2}{\it \Psi}\left(H(F)\right)\right.\\
\left.\mbox{for~infinitely~many}~F\in\mathcal{P}(\mathcal{M})\right\}.
\end{multline}
We denote the $k$-dimensional Lebesgue measure of $\mathcal{A}_k({\it\Psi,\mathcal{M}})$ by $|\mathcal{A}_k({\it\Psi,\mathcal{M}})|$. The following theorem is originally proved for non-degenerate manifolds in~\cite{BBKM02}. Here, we give the special case using $\mathcal{P}(\mathcal{M})$, where $\mathcal{M}$ is a set of linearly independent analytic functions~\cite{B01,BBKM02,Waldshmidt11}.
\begin{Theorem}[\cite{BBKM02}]\label{th:metric0}
Let ${\it \Psi}$ be a decreasing function and $\mathcal{M}$ be a set of linearly independent analytic functions, then
\begin{equation}~\label{eq:metricalresult}
|\mathcal{A}_k({\it\Psi},\mathcal{M})|=
\left\{\begin{array}{ll}
0,&\mbox{if}~\sum_{h=1}^\infty{\it\Psi}(h)<\infty,\\
1,&\mbox{if}~\sum_{h=1}^\infty{\it\Psi}(h)=\infty.\\
\end{array}
\right.
\end{equation}
\end{Theorem}
Note that $\mathcal{A}_k({\it\Psi,\mathcal{M}})$ is similar to $\mathcal{A}(\mathcal{M})$ because there is a close relation between polynomial approximation and approximating by algebraic numbers and hypersurfaces, see Section 2.1 of~\cite{Waldshmidt11}. The above uniform result has an asymptotic counterpart.
\begin{Theorem}[\cite{schmidt07}]\label{th:metric}
Let ${\it\Psi}$ be a decreasing function such that $\sum_{h=1}^\infty{\it\Psi}(h)$ diverges and $\mathcal{M}$ be a set of linearly independent analytic functions. Then for almost all ${\bf y}=(y_1,\ldots,y_k)\in[0,1]^k$,
\begin{equation}~\label{eq:metric}
\delta_\infty({\bf y},\mathcal{A}(F))<H(F)^{(-m+1)}{\it\Psi}(H(F))
\end{equation}
has infinitely many solutions $F\in\mathcal{P}(\mathcal{M})$.
\end{Theorem}
\begin{Example}
Let $\mathcal{M}=\{1,x\}$ and $m=2$, then the above theorem reduces to the Theorem~\ref{th:ADT}. In particular, infinitely many $\frac{p}{q}\in\mathbb{Q}$ in the Asymptotic Dirichlet's Theorem produce infinitely many univariate functions $F(x)=qx-p$ with $H(F)=q$. In this case, the approximation error decays with exponent $2$, that is $H(F)^{-2}=q^{-2}$.
\end{Example}
\begin{Example}
Replacing $k=1$ and $\mathcal{M}=\{1,x,x^2\}$ with $m=3$ in Theorem~\ref{th:metric}, the above theorem is the metric result of the famous Davenport and Schmidt theorem in~\cite{DS67}. For this result the approximation error is of order $H(F)^{-3}$, where the function $F$'s are quadratic equations.
\end{Example}
%%%%%%%%%%%%%%%%%%%%%%%%%%%%%%%%%%%%%%%%%%%%%%%%%%%%%%%%%%%%%%%%%%%%%%%%%%%%%%%%%%%%%%%%%%%%%%%%%%%%%%%%%
\section{Proof of Theorem~\ref{th:rateprecoder}}~\label{app:thrate}
To prove Theorem~\ref{th:rateprecoder}, we start from Lemma~\ref{lem:fixedaphi} and~\ref{lem:fixeda}.
\begin{IEEEproof}[\bf Proof of Lemma~\ref{lem:fixedaphi}]
Setting $\partial{Q'}/\partial{\alpha^H}=0$ with $Q'$ defined in \eqref{firstminimization}, we obtain
$$\rho\alpha\|{\bf h}\|^2-\rho{\bf a}\Phi^H{\bf h}^H+\alpha=0,$$
which implies that $\alpha_{\tiny\mbox{opt}}'=\frac{\rho{\bf a}\Phi^H{\bf h}^H}{1+\rho\|{\bf h}\|^2}$.
\end{IEEEproof}
\begin{IEEEproof}[\bf Proof of Lemma~\ref{lem:fixeda}]
We prove this lemma for $L=2$. The proof for $L>2$ is similar to this case and we omit it for the sake of brevity. We find
$${\bf h}'=(h_1',h_2')=\left(h_{1}e^{i\phi_1},h_{2}e^{i\phi_2}\right)=\left(\eta_1e^{i(\theta_1+\phi_1)},\eta_2e^{i(\theta_2+\phi_2)}\right).$$
We need to maximize $\left|\langle{\bf h}',{\bf a}\rangle \right|^2$ to achieve the highest computation rate. We have that
\begin{eqnarray}~\label{UBlll}
\left|\langle{\bf h}',{\bf a}\rangle \right|^2&=&\left|h'_{1}a_{1}^H+h'_{2}a_{2}^H\right|^2=\left|\eta_1e^{i(\theta_1+\phi_1)}\beta_1e^{i(-\psi_1)}+\eta_2e^{i(\theta_2+\phi_2)}\beta_2e^{i(-\psi_2)}\right|^2\nonumber\\
&=&\left|\left(\eta_1\beta_1\cos(\theta_1+\phi_1-\psi_1)+\eta_2\beta_2\cos(\theta_2+\phi_2-\psi_2)\right.\right.\nonumber\\
&+&\left.\left.i(\eta_1\beta_1\sin(\theta_1+\phi_1-\psi_1)+\eta_2\beta_2\sin(\theta_2+\phi_2-\psi_2)\right)\right|^2\nonumber\\
&=&\eta_1^2\beta_1^2\!+\!2\eta_1\eta_2\beta_1\beta_2\cos(\theta_1+\phi_1-\psi_1-(\theta_2+\phi_2-\psi_2))+\eta_2^2\beta_2^2\nonumber,
\end{eqnarray}
which means that in order to maximize $\left|\langle{\bf h}',{\bf a}\rangle \right|^2$, the phases $\phi_1$ and $\phi_2$ have to satisfy
\begin{equation}~\label{eq:just}
\theta_1+\phi_1-\psi_1=\theta_2+\phi_2-\psi_2.
\end{equation}
Thus, we obtain $\phi_1^{\tiny\mbox{opt}}=\psi_1-\theta_1$ for the first transmitter and $\phi_2^{\tiny\mbox{opt}}=\psi_2-\theta_2$ for the second transmitter.
The computation rate without phase precoder is
$$\mathfrak{R}(\rho,{\bf h},{\bf a})=\log^+\left(1+\rho\|{\bf h}\|^2\right)-\log^+\left(\|{\bf a}\|^2+\rho\left(\|{\bf h}\|^2\|{\bf a}\|^2 -\left|\langle{\bf h},{\bf a}\rangle\right|^2\right)\right).$$
If we use phases $\phi_\ell=\theta_\ell-\psi_\ell$, for $1\leq \ell\leq L$, then we get
$\left|\langle{\bf h}',{\bf a}\rangle \right|^2= \eta_1^2\beta_1^2+2\eta_1\eta_2\beta_1\beta_2+\eta_2^2\beta_2^2$.
On the other hand,
\begin{eqnarray}~\label{UBlll1}
\left|\langle{\bf h},{\bf a}\rangle \right|^2&=& \left|h_{1}a_{1}^H+h_{2}a_{2}^H\right|^2= \left|\eta_1e^{i(\theta_1)}\beta_1e^{i(-\psi_1)}+\eta_2e^{i(\theta_2)}\beta_2e^{i(-\psi_2)}\right|^2\nonumber\\
&=& \left|\left(\eta_1\beta_1\cos(\theta_1-\psi_1)+\eta_2\beta_2\cos(\theta_2-\psi_2) \right.\right.\nonumber\\
&+& \left.\left.i(\eta_1\beta_1\sin(\theta_1-\psi_1)+\eta_2\beta_2\sin(\theta_2-\psi_2))\right)\right|^2\nonumber\\
&=& \eta_1^2\beta_1^2+2\eta_1\eta_2\beta_1\beta_2\cos(\theta_1-\psi_1+\theta_2-\psi_2)+\eta_2^2\beta_2^2\nonumber.
\end{eqnarray}
It is clear that $\left|\langle{\bf h}',{\bf a}\rangle \right|^2\geq \left|\langle{\bf h},{\bf a}\rangle \right|^2$, which implies
that $\mathfrak{R}'(\rho,{\bf h},\Phi^{\tiny\mbox{opt}},{\bf a})\geq \mathfrak{R}(\rho,{\bf h},{\bf a})$.
The proof of Theorem~\ref{th:rateprecoder} is now obvious from the proof of the above two lemmas.
\end{IEEEproof}
With the above two Lemmas, the proof of Theorem~\ref{th:rateprecoder} is clear.
%%%%%%%%%%%%%%%%%%%%%%%%%%%%%%%%%%%%%%%%%%%%%%%%%%%%%%%%%%%%%%%%%%%%%%%%%%%%%%%%%%%%%%%%%%%%%%%%%%%%%%%%%%%%%%%%%%%%%
\section{Proof of Theorem~\ref{th:DoF}}~\label{app:thDoF}
%%%%%%%%%%%%%%%%%%%%%%%%%%%%%%%%%%%%%%%%%%%%%%%%%%%%%%%%%%%%%%%%%%%%%%%%%%%%%%%%%%%%%%%%%%%%%%%%%%%%%%%%%%%%%%%%%%%%%
To prove Theorem~\ref{th:DoF} we need the following lemma.
\begin{lemma}~\label{lem:huge}
For vectors ${\bf h}\in\mathbb{R}^{2L}$ and ${\bf w}'\in\mathbb{R}^{2L}$ obtained from ${\bf w}\in\mathbb{Z}^{2L}$ by ${\bf w}'={\bf w}\tilde{\Phi}^T$, where $\tilde{\Phi}_{2L\times 2L}$ is an orthogonal matrix and $L>1$, we have:
\begin{equation}\label{ineq}
\|{\bf w}'\|^2\left\|\frac{{\bf h}}{\|{\bf h}\|}-\frac{{\bf w}'}{\|{\bf w}'\|}\right\|^2\leq\left\|\frac{{\bf h}}{\|{\bf h}\|}-{\bf w}'\right\|^2,
\end{equation}
if $\left\|\frac{{\bf h}}{\|{\bf h}\|}-{\bf w}'\right\|_\infty\leq c^{-(L+1)}$, for a positive integer constant $c\geq2$.
\end{lemma}
\begin{IEEEproof}
If $\|{\bf w}'\|=1$, the inequality \eqref{ineq} becomes equality. We let $\widehat{\bf h}=\frac{{\bf h}}{\|{\bf h}\|}$ and we note that $\|\widehat{\bf h}\|=1$. The inequality \eqref{ineq} is equivalent to the following set of inequalities:
\begin{align*}
&         &  \|(\|{\bf w}'\|)\widehat{\bf h}-{\bf w}'\|^2&\leq\|\widehat{\bf h}-{\bf w}'\|^2 \\
&\Longleftrightarrow         &  \|\widehat{\bf h}\|^2\|{\bf w}'\|^2-2\|{\bf w}'\|\left\langle\widehat{\bf h},{\bf w}'\right\rangle+\|{\bf w}'\|^2&\leq\|\widehat{\bf h}\|^2-2\left\langle\widehat{\bf h},{\bf w}\right\rangle+\|{\bf w}'\|^2\\
&\Longleftrightarrow   &  \|{\bf w}'\|^2-2\|{\bf w}'\|\left\langle\widehat{\bf h},{\bf w}'\right\rangle&\leq1-2\left\langle\widehat{\bf h},{\bf w}'\right\rangle\\
&\Longleftrightarrow   &  2\left\langle\widehat{\bf h},{\bf w}'\right\rangle-2\|{\bf w}'\|\left\langle\widehat{\bf h},{\bf w}'\right\rangle&\leq 1-\|{\bf w}'\|^2\\
&\Longleftrightarrow   &  2\left\langle\widehat{\bf h},{\bf w}'\right\rangle(1-\|{\bf w}'\|)&\leq(1+\|{\bf w}'\|)(1-\|{\bf w}'\|)
\end{align*}
%$$\|(\|{\bf w}'\|)\widehat{\bf h}-{\bf w}'\|^2\leq\|\widehat{\bf h}-{\bf w}'\|^2$$
%$$\Longleftrightarrow\|\widehat{\bf h}\|^2\|{\bf w}'\|^2-2\|{\bf w}'\|\left\langle\widehat{\bf h},{\bf w}'\right\rangle+\|{\bf w}'\|^2\leq\|\widehat{\bf h}\|^2-2\left\langle\widehat{\bf h},{\bf w}\right\rangle+\|{\bf w}'\|^2$$
%$$\Longleftrightarrow\|{\bf w}'\|^2-2\|{\bf w}'\|\left\langle\widehat{\bf h},{\bf w}'\right\rangle\leq1-2\left\langle\widehat{\bf h},{\bf w}'\right\rangle$$
%$$\Longleftrightarrow2\left\langle\widehat{\bf h},{\bf w}'\right\rangle-2\|{\bf w}'\|\left\langle\widehat{\bf h},{\bf w}'\right\rangle\leq 1-\|{\bf w}'\|^2$$
%$$\Longleftrightarrow2\left\langle\widehat{\bf h},{\bf w}'\right\rangle(1-\|{\bf w}'\|)\leq(1+\|{\bf w}'\|)(1-\|{\bf w}'\|).$$
We now distinguish between two cases. Let $1-\|{\bf w}'\|>0$, we show that $2\left\langle\widehat{\bf h},{\bf w}'\right\rangle\leq1+\|{\bf w}'\|$.
In a triangle with sides ${\bf w}'$, $\widehat{\bf h}$, and $\widehat{\bf h}-{\bf w}'$, the law of cosines implies that
\begin{eqnarray}
2\left\langle\widehat{\bf h},{\bf w}'\right\rangle&=&\|{\bf w}'\|^2+ \|\widehat{\bf h}\|^2- \|\widehat{\bf h}-{\bf w}'\|^2=\|{\bf w}'\|^2+ 1-\|\widehat{\bf h}-{\bf w}'\|^2\nonumber\\
&\leq&\|{\bf w}'\|^2+1-\left(1-\|{\bf w}'\|\right)^2\label{eq:lemhu}\\
&=&2\|{\bf w}'\|<1+\|{\bf w}'\|,\nonumber
\end{eqnarray}
where \eqref{eq:lemhu} is obtained from triangle inequality $\left|\|\widehat{\bf h}\|-\|{\bf w}'\|\right|\leq\|\widehat{\bf h}-{\bf w}'\|$ and the last inequality is derived from the assumption $\|{\bf w}'\|<1$.
On the other hand, let $1-\|{\bf w}'\|<0$, then we need to show that $2\left\langle\widehat{\bf h},{\bf w}'\right\rangle\geq1+\|{\bf w}'\|$.
Again by using the triangle with sides ${\bf w}'$, $\widehat{\bf h}$, and $\widehat{\bf h}-{\bf w}'$, and the law of cosines, we have
\begin{eqnarray}
2\left\langle\widehat{\bf h},{\bf w}'\right\rangle&=&\|{\bf w}'\|^2+ \|\widehat{\bf h}\|^2- \|\widehat{\bf h}-{\bf w}'\|^2=\|{\bf w}'\|^2+ 1-\|\widehat{\bf h}-{\bf w}'\|^2\nonumber\\
&\geq&\|{\bf w}'\|^2+ 1-2L\|\widehat{\bf h}-{\bf w}'\|_\infty^2\label{eq:norm2toinfty}\\
&\geq&\|{\bf w}'\|^2+1-2Lc^{-2(L+1)}\label{eq:lemhu1}\\
&\geq&\|{\bf w}\|^2+1-1=\|{\bf w}'\|^2\label{eq:lemhu2}\\
&\geq&1+\|{\bf w}'\|,\nonumber
\end{eqnarray}
where \eqref{eq:norm2toinfty} is true since for a $2L$-dimensional vector ${\bf v}$ we have $\|{\bf v}\|^2\leq2L\|{\bf v}\|_\infty^2$, \eqref{eq:lemhu1} is true because of our condition in the statement of the Lemma, \eqref{eq:lemhu2} is correct since $c\geq2$, and the last inequality follows due to the fact that $\|{\bf w}'\|=\|{\bf w}\|$ is the norm of a non-zero integer vector. This completes the proof of Lemma~\ref{lem:huge}.
\end{IEEEproof}
Now, we proceed to proof the main result in this paper.
\begin{IEEEproof}[\bf Proof of Theorem~\ref{th:DoF}]
Let us first investigate ${\bf a}'={\bf a}\tilde{\Phi}^T$, where ${\bf a}=(a_1,a_2,\ldots,a_{2L})\in\mathbb{Z}^{2L}$ and
$$\tilde{\Phi}=
\left(
\begin{array}{cccccc}
\cos\phi_1&\cdots&0&\sin\phi_1&\cdots&0\\
\vdots&\ddots&\vdots&\vdots&\ddots&\vdots\\
0&\cdots&\cos\phi_L&0&\cdots&\sin\phi_L\\
-\sin\phi_1&\cdots&0&\cos\phi_1&\cdots&0\\
\vdots&\ddots&\vdots&\vdots&\ddots&\vdots\\
0&\cdots&-\sin\phi_L&0&\cdots&\cos\phi_L
\end{array}
\right).$$
One can rewrite ${\bf a}'$ as
$$(a_1\cos(\phi_1)+a_{L+1}\sin(\phi_{1}),a_2\cos(\phi_2)+a_{L+2}\sin(\phi_{2}),\ldots,-a_{L}\sin(\phi_{L})+a_{2L}\cos(\phi_{L})).$$
Next, we find a polynomial $F'$ with ${\bf a}'\in\mathcal{A}(F')$. It is clear that for every $1\leq \ell\leq L$, we have
$$\left(a'_{\ell}\right)^2+\left(a'_{\ell+L}\right)^2=\left(a_{\ell}\right)^2+\left(a_{\ell+L}\right)^2\in\mathbb{Z}.$$
By selecting coefficients $c_1,\ldots,c_L\in\mathbb{Z}$, we can form a polynomial $F'$ in variables $x_1,\ldots,x_{2L}$, which satisfies ${\bf a}'\in\mathcal{A}(F')$. More specifically, let us assume that we have chosen $c_1,\ldots,c_L\in\mathbb{Z}$, then we set
$$c_{L+1}=-\left(c_1\left(a_1^2+a_{L+1}^2\right)+\cdots+c_{L}\left(a_{L}^2+a_{2L}^2\right)\right).$$
Hence, $F'$ with
$$F'(x_1,\ldots,x_{2L})=c_1\left(x_1^2+x_{L+1}^2\right)+\cdots+c_{L}\left(x_{L}^2+x_{2L}^2\right)+c_{L+1}.$$
admits a zero at ${\bf a}'$. Hence, the vectors ${\bf a}'$ of the form ${\bf a}'={\bf a}\tilde{\Phi}^T$ are all roots of multivariate functions $F'$ formed by integer linear combinations of functions in $\mathcal{M}$ given in Example~\ref{ex:Mlinindanalytics}. Let us investigate the set of zeros of a function $F'\in\mathcal{P}(\mathcal{M})$, where $\mathcal{M}$ is as in Example~\ref{ex:Mlinindanalytics}. Since the domains of $G_\ell$, for $1\leq \ell \leq L$ are the closed circles around origin with different radiuses $r=a_\ell^2+a_{\ell+L}^2$, for $a_\ell,a_{\ell+L}\in\mathbb{Z}$, every solution ${\bf a}'$ to $F'=0$ is in the form of ${\bf a}'={\bf a}\tilde{\Phi}^T$ for some phases $(\phi_1,\ldots,\phi_{L})$. In essence we quantize the channel coefficients with the points on the circles of radius $r$ rather than the Gaussian integers only. Next, we use Theorem~\ref{th:metric} with $\mathcal{M}$ of Example~\ref{ex:Mlinindanalytics} to complete the proof. For a given ${\bf a}$, we have
\begin{eqnarray}
\mbox{PP Loss Term}&=&\min_{\substack{{\bf a}\in\mathcal{A}(F)\\F\in\mathcal{P}(\mathcal{M})}}\|{\bf a}\|^2+\rho\left(\|{\bf h}'\|^2\|{\bf a}\|^2 -\left|\langle{\bf h}',{\bf a}\rangle\right|^2\right)\nonumber\\
&=&\min_{\substack{{\bf a}\in\mathcal{A}(F)\\F\in\mathcal{P}(\mathcal{M})}}\|{\bf a}\|^2+\rho\|{\bf h}'\|^2\|{\bf a}\|^2\left(1-\cos^2(\angle({\bf h}',{\bf a}))\right)\nonumber\\
&=&\min_{\substack{{\bf a}\in\mathcal{A}(F)\\F\in\mathcal{P}(\mathcal{M})}}\|{\bf a}\|^2+\rho\|{\bf h}'\|^2\|{\bf a}\|^2\left(\sin^2(\angle({\bf h}',{\bf a}))\right)\nonumber\\
%&\leq&\min_{\substack{{\bf a}\in\mathcal{A}(F)\\F\in\mathcal{P}(\mathcal{M})}}\|{\bf a}\|^2+\rho\|{\bf h}'\|^2\|{\bf a}\|^2\angle({\bf h}',{\bf a})^2\nonumber\\
&\leq&\min_{\substack{{\bf a}\in\mathcal{A}(F)\\F\in\mathcal{P}(\mathcal{M})}}\|{\bf a}\|^2+\rho\|{\bf h}'\|^2\|{\bf a}\|^2\left\|\frac{{\bf h}'}{\|{\bf h}'\|}-\frac{{\bf a}}{\|{\bf a}\|}\right\|^2\label{eq:angleunits}
%&\leq&\|{\bf a}\|^2+4\rho\|{\bf h}\|^2\left\|\frac{{\bf h}}{\|{\bf h}\|}-{\bf a}\right\|^2\triangleq U({\bf a})\label{eq:lemma},\nonumber
\end{eqnarray}
where \eqref{eq:angleunits} is true because in an isosceles triangle with two equal sides $\frac{{\bf h}'}{\|{\bf h}'\|}$ and $\frac{{\bf a}}{\|{\bf a}\|}$ and angle $\angle({\bf h}',{\bf a})$ between these two sides, the length of the third side, which is $\left\|\frac{{\bf h}'}{\|{\bf h}'\|}-\frac{{\bf a}}{\|{\bf a}\|}\right\|$, is greater than or equal to the length of the perpendicular drawn into one of the equal sides, which is $\sin(\angle({\bf h}',{\bf a}))$. Theorem \ref{th:metric} implies that for ${\bf y}=\frac{{\bf h}'}{\|{\bf h}'\|}$, $k=2L$, ${\it \Psi}(h)=\frac{1}{h}$, and $\mathcal{M}$ as in Example~\ref{ex:Mlinindanalytics}, there are infinitely many polynomials $F'({\bf x})\in\mathbb{Z}[{\bf x}]\cap\mathcal{P}(\mathcal{M})$ each admits at least a root ${\bf a}'\in\mathcal{A}(F')$ such that
\begin{equation}~\label{metrictobeused}
\left\|\frac{{\bf h}'}{\|{\bf h}'\|}-{\bf a}'\right\|_\infty<H(F')^{(-(L+1)+1)}{\it \Psi}(H(F'))=H(F')^{-(L+1)}.
\end{equation}
Note that ${\bf a}'$ is now a solution to $F'\in\mathcal{P}(\mathcal{M})$ and has the form ${\bf a}'={\bf a}\tilde{\Phi}^T$ for ${\bf a}\in\mathbb{Z}^{2L}$ and some block phase matrix $\tilde{\Phi}^T$. We note that the author of~\cite{schmidt07} also investigates the ``size'' of such elements and have shown that the set of elements satisfying the above inequality have reasonably large volume. Therefore, we choose one $F'$ with $1\neq H(F')\geq2$. Hence, we get
\begin{eqnarray}
\min_{\substack{{\bf a}\in\mathcal{A}(F)\\F\in\mathcal{P}(\mathcal{M})}}\|{\bf a}\|^2+\rho\|{\bf h}'\|^2\|{\bf a}\|^2\left\|\frac{{\bf h}'}{\|{\bf h}'\|}-\frac{{\bf a}}{\|{\bf a}\|}\right\|^2 &\leq& \|{\bf a}'\|^2+4\rho\|{\bf h}'\|^2\|{\bf a}'\|^2\left\|\frac{{\bf h}'}{\|{\bf h}'\|}-\frac{{\bf a}'}{\|{\bf a}'\|}\right\|^2\nonumber\\
&\leq&\|{\bf a}'\|^2+\rho\|{\bf h}'\|^2\left\|\frac{{\bf h}'}{\|{\bf h}'\|}-{\bf a}'\right\|^2\label{eq:ineq0}\\
&\leq&\|{\bf a}'\|^2+2L\rho\|{\bf h}'\|^2\left\|\frac{{\bf h}'}{\|{\bf h}'\|}-{\bf a}'\right\|_\infty^2\label{eq:ineq}
\end{eqnarray}
where \eqref{eq:ineq0} follows from Lemma~\ref{lem:huge} for $c=H(F')$ while the inequality \eqref{eq:ineq} is true since for a $(2L)$-dimensional vector ${\bf v}$ we have $\|{\bf v}\|^2\leq2L\|{\bf v}\|_\infty^2$. From \eqref{eq:angleunits}, \eqref{metrictobeused}, and \eqref{eq:ineq}, we further upper bound the PP Loss Term as
\begin{eqnarray}
\mbox{PP Loss Term}&\leq&\|{\bf a}'\|^2+2L\rho\|{\bf h}'\|^2\left\|\frac{{\bf h}'}{\|{\bf h}'\|}-{\bf a}'\right\|_\infty^2\nonumber\\
&\leq&\|{\bf a}'\|^2+2L\rho\|{\bf h}'\|^2H(F')^{-2(L+1)}\label{eq:the1term}.
\end{eqnarray}
Note that $F'({\bf a}')=0$. Based on Theorem~\ref{th:Lojasiewicz} for $\mathcal{K}=\{{\bf 0}\}$ and ${\bf v}={\bf 0}$, there exists ${\bf b}\in\mathcal{A}(F')$ which satisfies the following set of inequalities:
\begin{eqnarray}
\|{\bf b}\|^2&\leq&\left(c_1|F'({\bf 0})|\right)^{\frac{2}{c_2}}\leq\left(c_1H(F')\right)^{\frac{2}{c_2}}\label{eq:h}
\end{eqnarray}
where \eqref{eq:h} is obtained since $|F'({\bf 0})|$ is the absolute value of the constant term in $F'$ and it is obviously less than $H(F')$ by the definition of Height of $F$. Now we distinguish between two cases: ({\em i}) the $\|{\bf a}'\|^2$ does not satisfy \eqref{eq:h}, and ({\em ii}) it also satisfies \eqref{eq:h}. In the latter case $\|{\bf a}'\|^2\leq \left(c_1H(F')\right)^{\frac{2}{c_2}}$. In the former case, since $\|{\bf a}'\|$ is finite and $\left(c_1H(F')\right)^{\frac{2}{c_2}}\leq \|{\bf a}'\|^2$, there exists a positive constant $c_1'$ such that $\|{\bf a}'\|^2\leq \left(c_1'H(F')\right)^{\frac{2}{c_2}}$. Adding both the cases yields
\begin{equation}~\label{eq:Ofirstterm}
\|{\bf a}'\|^2=O\left(H(F')^{\frac{2}{c_2}}\right).
\end{equation}
Based on \eqref{eq:Ofirstterm}, it is now clear that the first term in \eqref{eq:the1term} is increasing and the second term is decreasing as functions of $H(F')$. From with \eqref{eq:the1term}, and \eqref{eq:Ofirstterm}, we further find an upper bound on the PP Loss Term
\begin{equation}
\mbox{PP Loss Term}\leq \Theta\left(H(F')^{\frac{2}{c_2}}\right)+2L\rho\|{\bf h}'\|^2H(F')^{-2(L+1)}.
\end{equation}
In order to tighten the above inequality, we should equalize these two terms as function of $H(F')$. It follows that
$$H(F')=\Theta\left(\rho^{(2(L+1)+2/c_2)^{-1}}\right),$$
and hence
$$\|{\bf a}'\|^2+2L\rho\|{\bf h}'\|^2H(F')^{-2(L+1)}\leq O\left(\rho^{\frac{2(2(L+1)+2/c_2)^{-1}}{c_2}}\right).$$
This implies that
$$\mathfrak{R}'(\rho,{\bf h},\tilde{\Phi},{\bf a}) \geq \frac{1}{2}\log\left(1+\rho\|{\bf h}'\|^2\right)-\frac{1}{2}\log\left(\rho^{\frac{2(2(L+1)+2/c_2)^{-1}}{c_2}}\right)$$
As $\rho\rightarrow\infty$, it turns out that
$$\limsup_{\rho\rightarrow\infty}\frac{\max_{{\bf a}\in \mathbb{Z}^{2L}}\mathfrak{R}'(\rho,{\bf h},\tilde{\Phi},{\bf a})}{\frac{1}{2}\log \rho}\geq\frac{2(L+1)}{2(L+1)+\frac{2}{c_2}}.$$
This completes the proof.
\end{IEEEproof}

\end{document}